\documentclass[trackchanges, twocolumn]{aastex7}

\shorttitle{GLASHES II}
\shortauthors{K. Morii et al.}
\usepackage{textcomp}
\usepackage{comment}
\usepackage{rotating}
\usepackage{multirow}
\usepackage{makecell}
\usepackage{subcaption}
\usepackage{amsmath}

\newcommand{\hcop}{HCO$^{+}$}
\newcommand{\hnc}{HNC}
\newcommand{\dcop}{DCO$^{+}$}
\newcommand{\ntdp}{N$_2$D$^{+}$}
\newcommand{\dcn}{DCN}

\begin{document}

\title{Global and Local Infall in the ASHES Sample (GLASHES). II. Asymmetric Line Profiles around Dense Cores in 70 \textmu m Dark Massive Clumps} 

\author[orcid=0000-0002-6752-6061,sname='Morii']{Kaho Morii}
\altaffiliation{CfA Postdoctoral Fellow}
\affiliation{Center for Astrophysics $|$ Harvard \& Smithsonian, 60 Garden Street, Cambridge, MA 02138, USA}
\email[show]{kaho.morii@cfa.harvard.edu}  

\author[orcid=0000-0002-7125-7685]{Patricio Sanhueza}
\affiliation{Department of Astronomy, School of Science, The University of Tokyo, 7-3-1 Hongo, Bunkyo-ku, Tokyo 113-0033, Japan}
\email{psanhueza@astron.s.u-tokyo.ac.jp} 

\author[orcid=0000-0002-7125-7685]{Qizhou Zhang}
\affiliation{Center for Astrophysics $|$ Harvard \& Smithsonian, 60 Garden Street, Cambridge, MA 02138, USA}
\email{qzhang@cfa.harvard.edu} 

\author[orcid=0000-0002-3466-6164]{James M. Jackson}
\affiliation{Green Bank Observatory, 155 Observatory Road, Green Bank, WV 24944, USA}
\affiliation{USRA SOFIA Science Center, NASA Ames Research Center, Moffett Field, CA 94045, USA}
\affiliation{School of Mathematical and Physical Sciences, University of Newcastle, University Drive, Callaghan, NSW 2308, Australia}
\email{jmjackso@nrao.edu} 

\begin{abstract}
Gravitational collapse is fundamental to star formation, yet direct kinematic evidence of infall at the core scale in high-mass star-forming regions remains poorly constrained. 
We present the first large-scale statistical study of infall signatures in 304 dense cores within 24 massive 70 $\mu$m-dark clumps from the GLASHES (Global and Local Infall in the ASHES Sample) survey. 
Using ALMA Band 6 observations of the optically thick tracers \hcop\ and HNC ($J$=3--2), we systematically characterize blue asymmetry line profiles indicative of infalling motions. 
We employ two complementary metrics, the velocity difference parameter ($\delta_v$) and the asymmetry parameter ($A$), to quantify infall signatures, finding consistent results across both tracers. 
Blue asymmetry profiles are detected in $\sim$50-60\% of cores ($\delta_v<$0 or $A>$0). 
Spectral classification reveals that $\sim$60\% of cores exhibit double-peaked profiles, and 34\% and 39\% show blue asymmetry profiles in \hcop\ and \hnc, respectively, with the percentage increasing with core mass and surface density. 
Accounting for geometric effects that can obscure infall signatures, our results suggest that gravitational collapse is prevalent in and around the cores. 
Importantly, infall signatures are detected from the prestellar stage and become more dominant as cores' evolution proceeds. 
Even cores with virial parameters $\alpha_{vir} > 2$ show infall signatures, suggesting that external compression may trigger collapse in addition to self-gravity or that linewidth may include inward motion in addition to turbulence. 
Furthermore, a moderate correlation between clump-scale and core-scale asymmetry supports a hierarchical collapse scenario, implying a dynamic and multi-scale process of high-mass star formation. 
\end{abstract}

\keywords{\uat{Infrared dark clouds}{787} --- \uat{Star formation}{1569} --- \uat{Star forming regions}{1565}}

\section{Introduction} \label{sec:intro} 

Gravitational collapse is the fundamental process driving the formation of stars. 
As comprehensively reviewed by \citet{Evans99} and \citet{Evans03}, identifying the kinematic signatures of this collapse is crucial for understanding the earliest stages of star formation.
Specifically, the ``blue asymmetry'' profile, a line profile with brighter emission on the blue side, observed in optically thick molecular lines toward central dense regions, provides direct spectroscopic evidence of infalling gas motions \citep[e.g.,][]{Jackson19}.
Such a profile is produced when the excitation temperature increases toward the center of the structure (e.g., cloud, clumps, or cores).
This signature has been observed for about five decades across a wide range of environments, including molecular clouds \citep{LeungBrown77, Myers96}, dense cores in nearby low-mass star-forming regions \citep{Zhou93, Tafalla98, LeeMyers11, 
Campbell16}, dense cores in high-mass star-forming regions \citep{Csengeri11, Contreras18, Morii25, Gupta26}, high-mass protostellar cores \citep{HoHaschick86, KetoHoHaschick88, ZhangHoOhashi98, FullerWilliamsSridharan05}, massive clumps \citep{Sanhueza10,Schneider10, 
Reiter11, Rygl13, He16, Wyrowski16, Traficante17, Jackson19, Yang21, Xu23_hcn, Jackson26}, and (UC)H$\textsc{II}$ regions \citep{ZhangHo97}.
However, while infall has been well characterized at the core scale ($\sim$0.01--0.1~pc) in nearby low-mass star-forming regions, direct kinematic evidence at the same scale in high-mass star-forming regions remains limited to some case studies.

Understanding infall in high-mass star formation is of particular importance in the context of current theoretical frameworks.
Models such as the ``clump-fed'' accretion scenario \citep[e.g.,][]{Bonnell04, Wang10, Padoan20} including the Global Hierarchical Collapse (GHC) model \citep{vazquez19} predict that gas inflow or infall is continuous from the parent molecular cloud down to individual cores, implying that core-scale infall is a direct result of large-scale dynamical accretion. 
Characterizing infall at the core scale is therefore essential to validate these scenarios and to understand how high-mass stars accumulate sufficient mass. 
However, observational verification at the core scale in high-mass star-forming regions presents significant challenges due to their typically large distances ($\gtrsim 3$~kpc) and highly clustered environments, and previous studies have largely been limited to 
coarse spatial resolutions at the clump scale ($\sim$1~pc), or to a handful of particularly bright massive protostars and (UC)H\textsc{II} regions. 

At the clump scale, large surveys such as MALT90 \citep{Foster11,Jackson13} have successfully identified global infall motions \citep[e.g.,][]{Jackson19}, and suggest that infall is particularly active in earlier evolutionary stages, including infrared dark clouds \citep[IRDCs;][]{Menten05, Sridharan05, Rathborne06,Chambers09,Sanhueza12,Sanhueza13,Sanhueza17,Morii21}. 
Although some interferometric studies have inferred gas inflow from velocity gradients in IRDCs \citep[e.g.,][]{Csengeri11, Beuther13, Henshaw14, Lu18, Chen19, Sanhueza21, Redaelli22,Sanhueza25}, direct kinematic evidence of gravitational collapse via blue asymmetry at the core scale has been reported in only a few individual case studies \citep[e.g.,][]{Csengeri11, Contreras18, Morii25}.
A comprehensive statistical understanding of core-scale infall in the earliest stages of high-mass star formation therefore remains lacking. 

To address this observational gap, the ``Global and Local Infall in the ASHES Sample'' (GLASHES) survey targets 70~\textmu m dark massive clumps identified in the ALMA Survey of 70~\textmu m Dark High-mass Clumps in Early Stages (ASHES; \citealt{Sanhueza19,  Morii23}), which represent ideal laboratories for studying the initial conditions of high-mass star formation \citep[e.g.,][]{Li20,Tafoya21, Sakai21, Li22, Sabatini22, Li23, Izumi24, Morii24, Lin25,Morii26}. 
In our pilot study in one of the ASHES targets, G337.54, hosting 17 cores \citep[][]{Morii25}, we established a robust methodology to extract core-scale infall signatures, derived infall velocities, and confirmed that they are higher than those found in nearby low-mass star-forming regions.
Following it, this study extends the analysis to 24 massive clumps containing 304 dense cores to provide the first large-scale statistical evidence of gravitational collapse in the earliest stages of high-mass star formation.
The primary goal of this paper is to report the detection rates of blue asymmetry across a large sample and to rigorously evaluate the classification criteria using multiple metrics. 
We also investigate how the frequency of these signatures relates to core properties and the surrounding clump environment. 
Detailed radiative transfer modeling to derive quantitative infall rates and other physical parameters will be presented in a subsequent paper. 
We describe the observational setup in Section~\ref{sec:obs}, present the line emission results and methods for profile characterization and spectral classification in Section~\ref{sec:results}, discuss the detection rates and correlations with core and clump properties in Section~\ref{sec:discussion}, and summarize our findings in Section~\ref{sec:conclusion}.

\section{Observations and Data Reduction}
\label{sec:obs} 
The GLASHES project consists of data from Band 3 and 6 taken with the main 12-m array, and the Atacama Compact Array (ACA), including both the 7-m array and Total Power (TP) array. 
The observations have been done during Cycles 6 (2018.1.00299.S, PI: Y. Contreras), 10 (2023.1.01150.S, PI: K. Morii), and 11 (2024.1.01505.S, PI: K. Morii). 
Targets are 24 massive clumps selected from the ASHES sample, which contains at least one sub-virial ($\alpha_{\rm vir}<2$), intermediate-mass cores ($M>4\,M_\odot$). 
The correlator setup covered \hcop\ ($J=3-2$, $\nu$=267.557633 GHz), \hnc\ ($J=3-2$, $\nu$=271.981111 GHz) and HC$^{18}$O$^+$ ($J=3-2$), $\nu$=255.479 GHz in Band 6 and N$_2$H$^+$ ($J=1-0$, $\nu$=93.173763 GHz) in Band 3. 
In this paper, we mainly use Band 6 data. 
N$_2$H$^+$ data was used only for determining the systemic velocity (see Appendix~\ref{sec:Vsys}). 

The ALMA 12\,m array consisted of 41--47 antennas, with a baseline ranging from 15 to 313 m and 779 m for Band 6 and Band 3, respectively. 
The total on-source time was $\sim$7--20 minutes. 
These observations are sensitive to angular scales smaller than $\sim$10$''$ and $\sim$15$''$ for Band 6 and Band 3, respectively. 
More extended emission was recovered by including ACA data. 
The 7 m array observations consisted of 8--11 antennas, with baselines ranging from 5 to 48 m. 
The total on-source time was $\sim$50--97 minutes for Band 6 and $\sim$29--58 minutes for Band 3.   
These observations are sensitive to angular scales smaller than $\sim$25$''$ and $\sim$72$''$ for Band 6 and Band 3, respectively. 
We also obtained total power data to cover the zero-baseline information. 
The total power array consisted of three antennas, and the total on-source time was $\sim$300 minutes and $\sim$250 minutes for Band 6 and Band 3, respectively.   
All observations have been done in single-pointing. 
Such observation setup (observation dates, baseline, number of antennas, and total on-source time) is summarized in Table~\ref{tab:obs_set}. 

\begin{deluxetable*}{lcccccl}
\tablecaption{Summary of ALMA Observations\label{tab:obs_set}}
\tablehead{\colhead{Band} & \colhead{Config.} & \colhead{Obs. Dates} & \colhead{\makecell{Baselines\\(m)}} & \colhead{$N_{\rm ant}$} & \colhead{\makecell{Total time\tablenotemark{a}\\(min)}}
}
\startdata
\multirow{2}{*}{Band 6} & 12m (C43-1, C43-2) & 2024 Mar. 17--2024 Dec. 10 & 15.1--313.7 & 43--47 & 7--15 \\
 & 7m & 2024 Jan. 21--2024 Jun. 13 & 8.9--48.9 & 8--11 & 50--97 \\
\tableline
\multirow{2}{*}{Band 3} & 12m (C43-3, C43-4) & 2024 Jan. 15--2025 Jan. 25 & 15.1--783.5 & 35--46 & 13--22 \\
 & 7m & 2023 Oct. 25--2024 Jan. 26 & 8.9--48.9 & 7--10 & 29--58 \\
\tableline
\enddata
\tablenotetext{a}{On-source integration time per region.}
\end{deluxetable*}

Data reduction was carried out using CASA software package versions 6.5.4.9 and 6.6.1.17 for calibration and 6.6.3-22 for imaging \citep{CASA22}. 
After subtracting continuum emission, we combined the 12 m array data with the 7 m array data 
and cleaned together. 
After that, we combined TP data with the interferometric images using the CASA task, feather. 
We used TCLEAN with Briggs' robust weighting of 0.5 to the visibilities and an imaging option of MULTISCALE with scales of 0, 5, 15, 25, and 50 times the pixel size (0\farcs2), considering the spatially extended nature of the emission.  
We used the automasking algorithm, auto-multithresh \citep{Kepley19-automask}, where the parameters of sidelobethreshold, noisethreshold, lownoisethreshold, negativethreshold, and minbeamfrac are set to 1.75, 2, 1.5, 0, and 0.3, respectively, mostly following the recommended values by \citep{Kepley19-automask}. 

The achieved beam size of \hcop, \hnc\ and HC$^{18}$O$^+$ cubes are (0.89--1.06) arcsec $\times$ (1.05--1.29) arcsec, whilst N$_2$H$^+$ cubes have beam size of (1.10--1.38) arcsec $\times$ (1.40--1.80) arcsec. 
The velocity resolution of HNC, HCO$^+$ and HC$^{18}$O$^+$ is 0.27 km\,s$^{-1}$ and that of N$_2$H$^+$ is 0.19 km\,s$^{-1}$. 
Average 1$\sigma$ rms noise level measured from line-free channel is $\sim$7 and 5 mJy beam$^{-1}$, corresponding to 1.0 K and 0.3 K, for Band 6 and Band 3 cubes, respectively. 
All data shown in the paper about these three lines are ALMA 12\,m, {bf 7\,m}, and TP combined, and before the primary beam correction. 
Appendix~\ref{sec:TP-combination} discusses how combining TP data with interferometric data affects the line shape.

Additionally, in this paper, we also use the 1.3 mm continuum emission and the optically thin tracers of \dcop\ ($J$=3--2), \ntdp\ ($J$=3--2), \dcn\ ($J$=3--2) from the ASHES project \citep{Li22, Li23, Morii23, Morii24}, see Appendix~\ref{sec:Vsys}. 
These are 12m+7m combined datasets without TP combination, but with a similar beam size of $\sim$1.2''. 
Core physical information such as size, mass, density, virial parameter, and evolutionary stages has been taken from \citet{Morii23} and \citet{Morii24}.


\section{Results} \label{sec:results}
\subsection{Spatial distribution}
\begin{figure}
    \centering
    \includegraphics[width=0.95\linewidth]{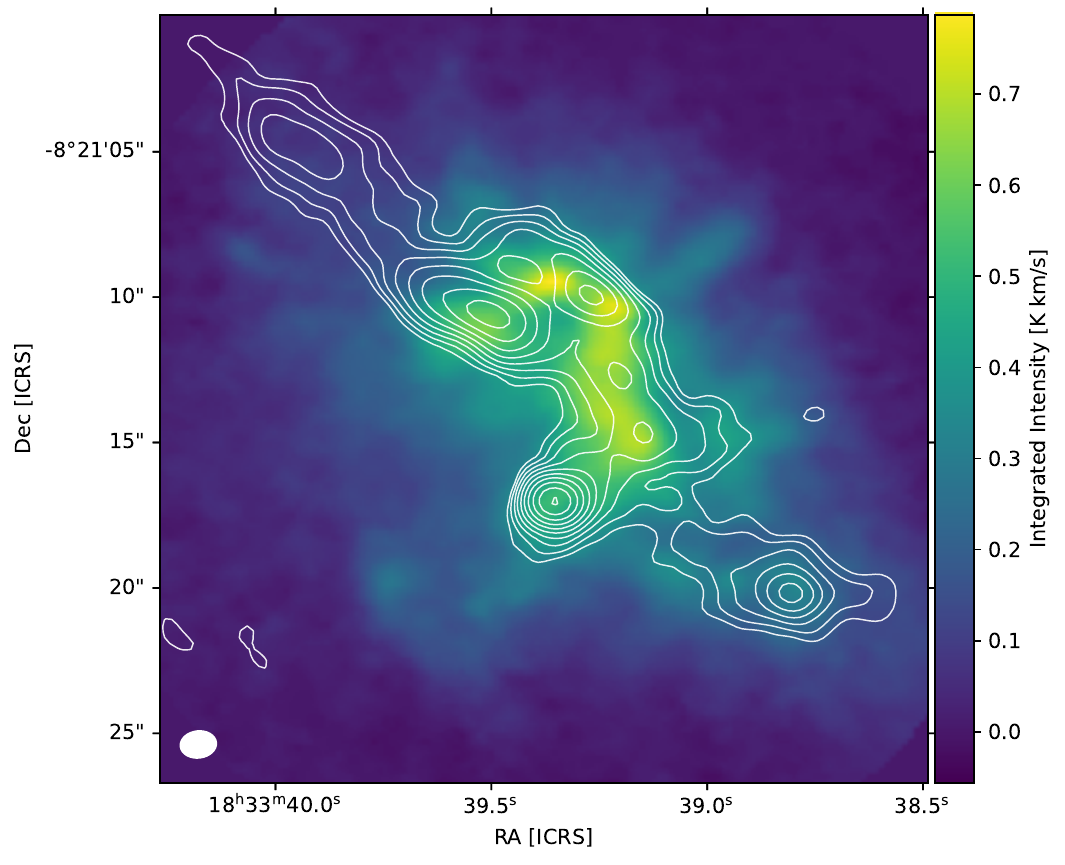}
    \caption{Moment 0 map of HNC overlaid with continuum emission as white contours. The beam size is shown in the bottom left corner. Contour levels are 5$\sigma$ $\times \sqrt{2}^n$ ($n=$1,2,3,...), where 1$\sigma$=0.1 mJy\,beam$^{-1}$. }
    \label{fig:mom0_cont}
\end{figure}

Figure~\ref{fig:mom0_cont} shows the integrated intensity (zeroth-order moment) map of HNC, overlaid with 1.3 mm dust continuum. It indicates the consistency between bright spots of HNC and dust continuum emission, as well as a more extended distribution of HNC emission than dust continuum. 

\begin{figure*}
    \centering
    \includegraphics[width=0.95\linewidth]{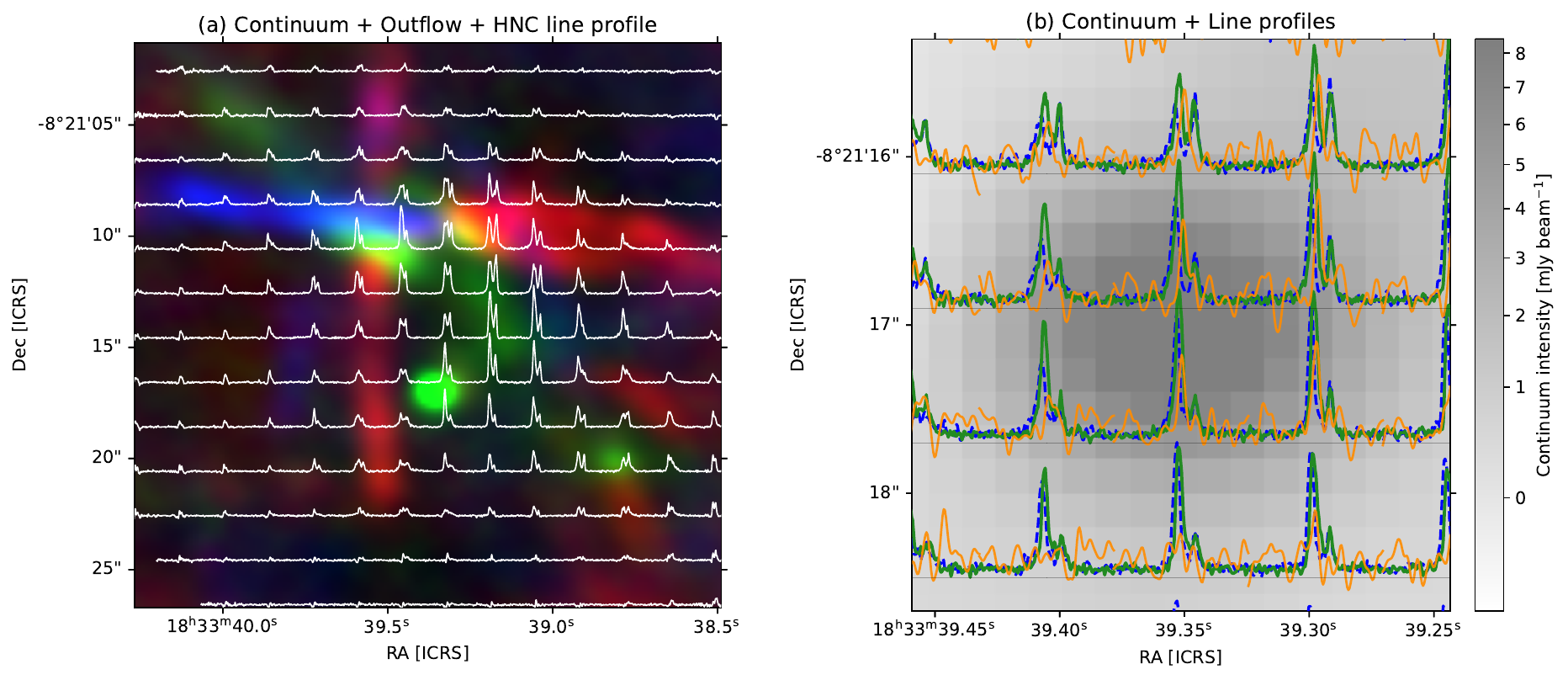}
    \caption{(a) HNC line-spectra map (binned to 10 pix $\times$ 10 pix) overlaid on a three-color composite image: CO outflow in red and blue, and continuum emission in green. (b) Line profiles of DCO$^+$ (orange), HCO$^+$ (blue dashed), and HNC (green) toward ALMA 1 (the brightest core), binned to 4 pix $\times$ 4 pix $\sim$1 beam. Gray-scale background is 1.3 mm continuum. Each spectrum is normalized by the brightest peak intensity.}
    \label{fig:cont_profile}
\end{figure*}

Figure~\ref{fig:cont_profile} shows (a) line spectra map of HNC overlaid on a three color image (r/b/g: CO redshifted, CO blue shifted, and continuum), and (c) zoom-in spectral map of HNC (green), HCO$^+$ (blue) and DCO$^+$ (orange) around the brightest core. 
It indicates that double peak profiles are seen widely spread in the field, not only around or inside dense cores but also in inter-core regions. Spectra that overlap with CO outflows sometimes have high-velocity or tail/wing components. 
Panel (b) shows the comparison of the spectra. HNC and HCO$^+$ show double peak profiles around the most massive core, where \dcop\ typically peaks at their dip velocity. 
The profiles of HNC and HCO$^+$ look generally the same, but sometimes differ in intensity and peak velocity, for example the profile at RA=18h33m39.40s and Decl.=--8d21m16s. 

\subsection{Line Detection}\label{sec:line_detection}
Line profiles of \hcop\ and \hnc\ toward the 304 identified cores were extracted by averaging within the cores. This was done by creating an ellipse mask from the major and minor axes and position angle information obtained from dendrogram analysis applied to continuum emission \citep[][]{Morii23}. 
HCO$^+$\ emission was detected in 282 cores, and HNC\ in 285 cores, at the $>5\sigma$ noise level. 
The detected profiles exhibit a variety of morphologies, including single peaks, skewed profiles, and double peaks with either the blue or red peak stronger. 
High-velocity line wings and tails indicative of outflows were also observed in a subset of cores, especially in \hcop. 
Figure~\ref{fig:class_profile} in Appendix~\ref{sec:profile-classification} presents several examples of the extracted profiles. 

Using the HC$^{18}$O$^+$ ($J=3-2$) as an optically thin reference, we estimated the peak optical depth of HCO$^+$ ($J=3-2$) for 30 cores in three representative regions (see Appendix~\ref{sec:opticaldepth} for details). For cores where HC$^{18}$O$^+$ is detected, the derived optical depths ($\tau = 1.7-11.4$).
These values serve as lower limits due to the uncertainties in excitation temperature, thereby confirming that the optically thick assumption is valid for these sources. 
For cores without HC$^{18}$O$^+$ detection, predominantly low-mass prestellar cores, only upper limits could be obtained. However, none of these upper limits indicate clearly optically thin conditions. Therefore, in the following analysis, we treat both \hcop\ and HNC as optically thick tracers. 

\subsection{Infall Statistics}
Blue asymmetry profiles sometimes refer to both blue skew profiles and double peak profiles with the blue peak stronger (or blue double peak profile). 
To avoid contamination from the multi-velocity components in a line of sight, the blue asymmetry profile is confirmed with an optically thin line. 
If the optically thick line shows a double peak profile with a dip close to the systemic velocity, where an optically thin line shows a single peak, the profile is considered to be an indicator of inward or outward motion. If the blue-shifted component is brighter, it implies inward motion. 
A profile without a clear dip but with a blue skew, indicating that the peak is shifted to the blue side relative to the systemic velocity, is defined as a blue-skew profile. 
After determining the systemic velocity for all the cores (see Appendix~\ref{sec:Vsys}), we employed two methods to quantitatively characterize blue asymmetry profiles: 
\begin{enumerate}
    \item Velocity Difference by \citet{Mardones97}: 
    \begin{equation}
        \delta_v \equiv \frac{V_{\rm thick} - V_{\rm thin}}{\Delta V_{\rm thin}}
        \label{equ:delta_v}
    \end{equation}
    Here $V_{\rm thick}$ is the brightest peak velocity of the optically thick line, $V_{\rm thin}$ is the peak velocity of the optically thin line tracing the systemic velocity of the dense core, and $\Delta V_{\rm thin}$ is the full width at half maximum (FWHM) of the optically thin line. The blue asymmetries correspond to $\delta_v<0$. Restrictively, $\delta_v<-0.25$ is considered to be a significant blue asymmetric profile. 
    
    \item Asymmetry Parameter by \citet{Jackson19}: 
    \begin{equation}
       A \equiv \frac{I_{\rm blue} - I_{\rm red}}{I_{\rm blue} + I_{\rm red}},
       \label{equ:A}
    \end{equation}
    where $I_{\rm blue}$ is the integrated intensity at velocities less than the reference velocity and $I_{\rm red}$ is the integrated intensity at more positive velocities. 
    The integration range for $I_{\rm blue}$ extended from $V_{\rm thin} - 2\Delta V_{\rm thin}$ to $V_{\rm thin}$ and that for $I_{\rm red}$ was from $V_{\rm thin}$ to $V_{\rm thin} + 2\Delta V_{\rm thin}$, where $V_{\rm thin}$ is the reference line velocity and $\Delta V_{\rm thin}$ is the observed velocity FWHM of optically thin line.  
    $A$ represents the fraction of the total line flux that lies to the blueshifted (or redshifted) side of the systemic velocity. In this scheme, line profiles with a blue asymmetry will have values of $A>0$, symmetric line profiles will have $A=0$, and line profiles with a red asymmetry will have $A<0$. Larger absolute values of $A$ are more asymmetric line profile. 
    In the absence of absorption against a continuum source, $A$ is bounded between -1 and +1.
\end{enumerate}

\begin{figure}
    \centering
    \includegraphics[width=0.99\linewidth]{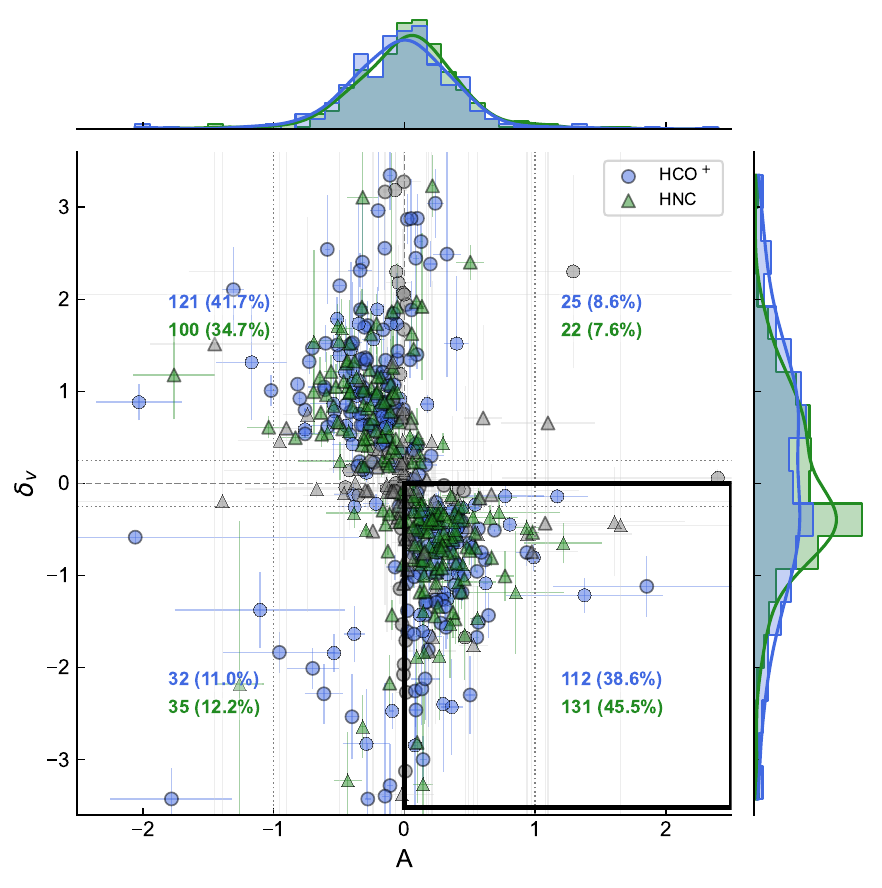}
    \caption{$\delta_v$ and $A$ scatter plots derived by HCO$^+$ (blue) and HNC line (green). Cores with large uncertainties ($>$100\%) are colored gray. The histogram and Kernel Density Estimation (KDE) plots for each parameter are shown on the right and top, respectively. The percentage of data points in each quadrant is denoted as colored text. The quadrant enclosed by a thick black line corresponds to the blue asymmetry in both parameters. Cores with a clear single peak detected either \dcop\ or N$_2$D$^+$ have black edge circles. Faint dotted lines mark $A=\pm 1$, beyond which ($|A|>1$) absorption is implied by definition, and $\delta_v\pm0.25$, the adopted velocity threshold (see main text).} 
    \label{fig:A_delta_scatterplot}
\end{figure}

Applying these methods to both \hcop\ and \hnc, we found almost consistent results across the two metrics and also across the lines. 
Figure~\ref{fig:A_delta_scatterplot} shows the $\delta_v$ and $A$ scatter plots derived for both \hcop\ (blue) and HNC (green). 
Cores with a large uncertainty ($>$100\%) are colored in gray. 
The uncertainties for both parameters are estimated following \citet{Jackson19}. 
The majority of the sample lies along the line from the top-left to the bottom-right, implying consistent classification from the $\delta_v$ and $A$ parameters. 
Quantitatively, in terms of $\delta_v$, 50.3\% and 58.0\% of cores have $\delta_v <$0, corresponding to inward motion, in HCO$^+$ and HNC, respectively. 
More restrictively, $\delta_v <$-0.25 are found in 40\% and 45\% of the whole sample. 
In terms of $A$, 48.3\% and 53.7\% of cores show $A>0$ in HCO$^+$ and HNC, respectively. 
Some cores show $|A|>1$ due to absorption in the profile. 
Compared to HCO$^+$, a higher percentage of inward-indicating results is obtained from HNC lines. This is clearly seen in the histogram of $\delta_v$, whose peak is located $<-0.25$. 
However, the fraction of blue asymmetry profiles is still not significantly higher than that of red asymmetry. 

\subsection{Line Profile} 
Beyond the quantitative asymmetry measures, we further characterize the profile by its shape. 
As discussed in Appendix~\ref{sec:TP-combination}, ignoring TP data may cause significantly different results in such analysis.
We measured the peak velocity and performed one- or two-Gaussian fits, and  
using the velocity dispersion ($\sigma$) and peak velocity separation ($\Delta V = |v_{\rm gauss, 1} - v_{\rm gauss, 2}|$), we classified spectrum into single, skew, double peak, or complex classes as summarized in Figure~\ref{fig:class-fig}.  
The detailed steps are described in Appendix~\ref{sec:profile-classification}.  
\begin{figure}
    \centering
    \includegraphics[width=0.95\linewidth]{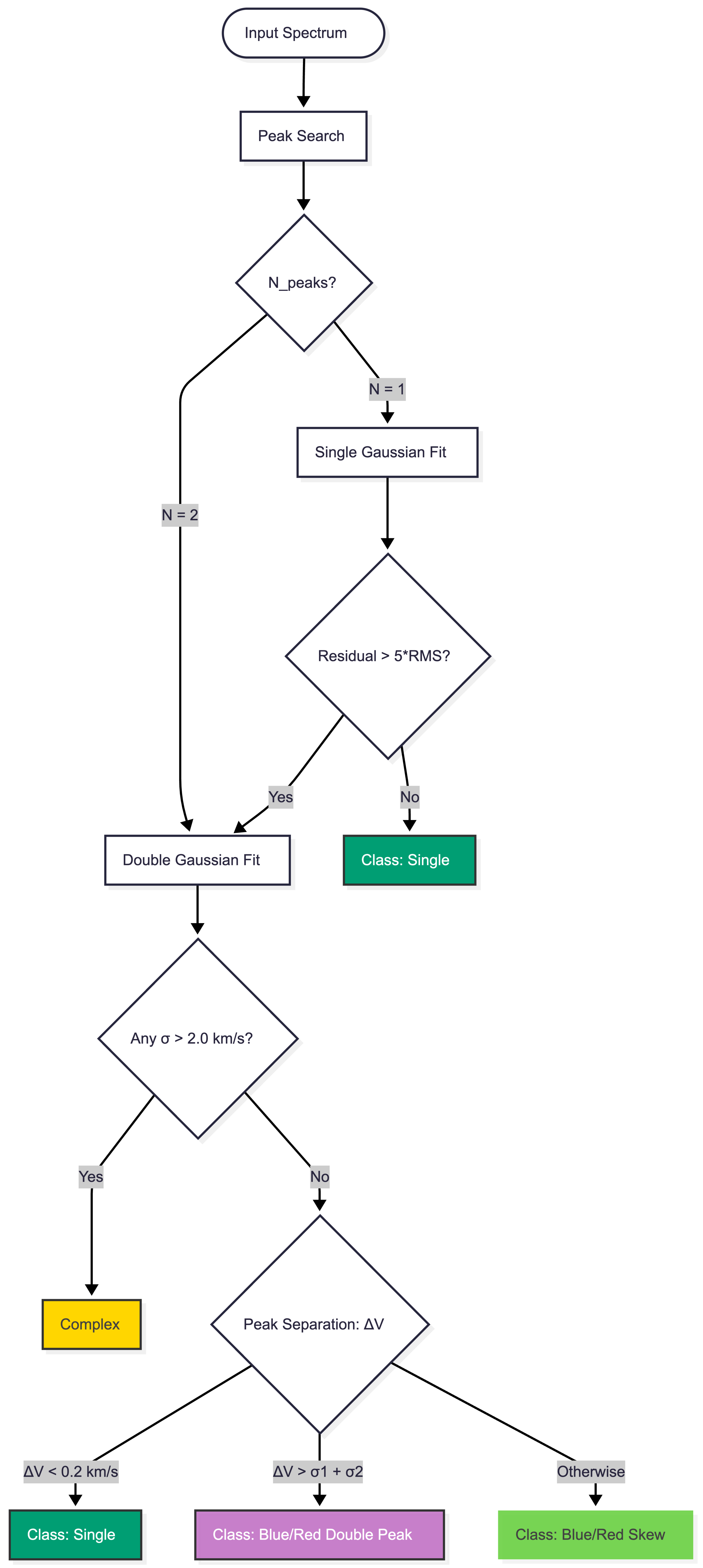}
    \caption{Decision tree for line profile classification. The algorithm determines the spectral class by sequentially evaluating the number of peaks, Gaussian fit results ($\sigma$ and $\Delta V$). Classes are color-coded: green for Single, purple for Double Peak, light green for Skew, and yellow for Complex.}
    \label{fig:class-fig}
\end{figure}

As a result, $\sim$60\% of the sample is classified as double peak profiles by both HCO$^+$ and HNC, which can be roughly described by two Gaussians. 
15\% show a skew profile without a clear dip or profile described by a closely located two Gaussians, 10\% show a single peak, and the other $\sim$15\% show a complex profile. 
Some examples of the classification are shown in Figure~\ref{fig:class_profile}. 
Figure~\ref{fig:class_corr} in the Appendix shows the correlation plot of line classification between HCO$^+$ and HNC, indicating the good correlation between the two lines, albeit with occasional discrepancies in classification. 
More detailed discussion of these classifications and cores' physical properties is presented in Section~\ref{sec:class-phy}. 
All the measured parameters and classifications are listed in Table~\ref{tab:infall}. 

\begin{deluxetable*}{lcccCcCCCCcc}
\tablecaption{Summary of Infall Parameters\tablenotemark{*} \label{tab:infall}}
\tablehead{
\colhead{Region} & \colhead{ID} & \colhead{Evolutionary} & \colhead{Gravitational} & \colhead{$V_{\rm cen}$} & \colhead{$\sigma_{\rm thin}$} & \multicolumn{2}{c}{$\delta$} & \multicolumn{2}{c}{$A$} & \multicolumn{2}{c}{Spectral Class} \\
 \colhead{} & \colhead{} & \colhead{Stage\tablenotemark{a}} & \colhead{Boundness\tablenotemark{b}} & \colhead{(km\,s$^{-1}$)} & \colhead{(km\,s$^{-1}$)} & \colhead{HCO$^+$} & \colhead{HNC} & \colhead{HCO$^+$} & \colhead{HNC} & \colhead{HCO$^+$} & \colhead{HNC}
}
\startdata
G016.97 & 1 & outflow & bound & 41.34 & 0.36 & -0.90 & -0.57 & 0.41 & 0.40 & blue double peak & blue double peak \\
G016.97 & 2 & warm & bound & 40.54 & 0.26 & 0.30 & 0.31 & 0.20 & 0.11 & red skew & red skew \\
G016.97 & 3 & prestellar & bound & 40.67 & 0.31 & -0.31 & -0.30 & 0.16 & 0.02 & blue double peak & blue skew \\
G016.97 & 4 & prestellar & non-detection & 40.41 & 0.34 & -0.30 & -0.28 & 0.41 & 0.21 & blue skew & blue skew \\
G016.97 & 5 & outflow & non-detection & 41.32 & 0.40 & -1.08 & -1.64 & 0.62 & 0.46 & blue double peak & blue double peak \\
G016.97 & 6 & prestellar & non-detection & 40.35 & 0.37 & -0.98 & -0.19 & 0.33 & 0.13 & complex & blue skew \\
G016.97 & 8 & prestellar & non-detection & 40.47 & 0.29 & 2.18 & -0.42 & -0.04 & -0.07 & red double peak & blue double peak \\
G016.97 & 9 & outflow & non-detection & 40.46 & 0.14 & -0.06 & -0.42 & 0.41 & 1.61 & single & single \\
G016.97 & 10 & prestellar & non-detection & 40.52 & 0.37 & 0.22 & 0.54 & -0.01 & -0.31 & red skew & red skew \\
G016.97 & 11 & prestellar & non-detection & 40.88 & 0.45 & 0.63 & -0.27 & -0.45 & -0.01 & red double peak & blue double peak 
\enddata
\tablenotetext{*}{Full version is available online as machine-readable table. }
\tablenotetext{a}{Determined based on the detections of CO outflow and warm gas tracers in \citet{Li22} and \citet{Morii24}.}
\tablenotetext{b}{Determined from the virial analysis considering gravity and non-thermal velocity dispersion excluding external pressure and magnetic field support \citep{Morii24}. }
\end{deluxetable*}

\begin{figure*}
    \centering
    \includegraphics[width=0.9\linewidth]{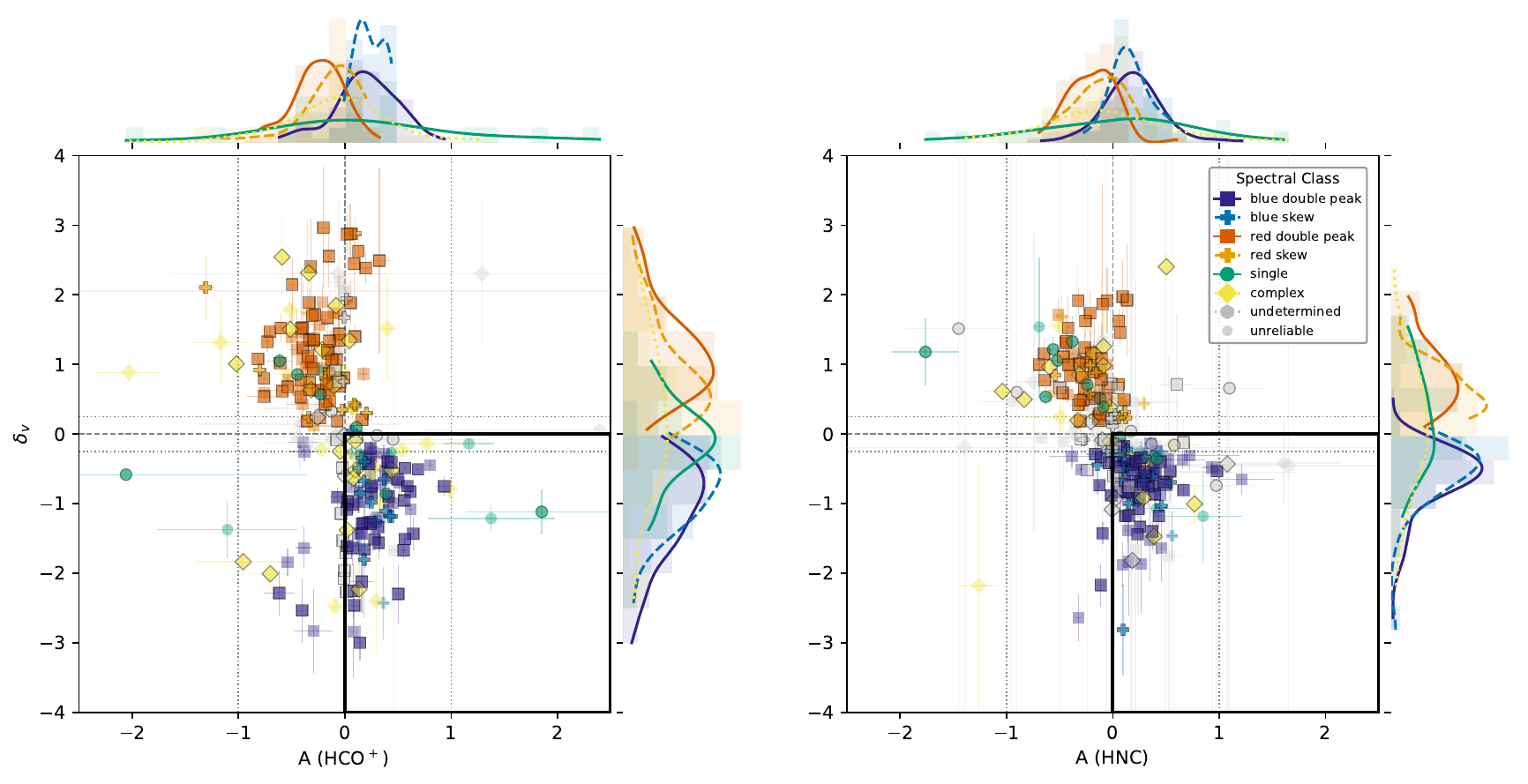}
    \caption{$\delta_v$ and $A$ scatter plots colored by classified categories. Solid and dashed KDE curves represent double-peak and skewed spectral profiles, respectively. As with Figure~\ref{fig:A_delta_scatterplot}, a black edge means the core has a clear single peak in \dcop\ or N$_2$D$^+$, while others have multiple peaks or no detections of \dcop\ nor N$_2$D$^+$. Cores with large uncertainties ($>$100\%) are colored gray.}
    \label{fig:Adelta_class}
\end{figure*}

How such classification correlates with $\delta_v$ and $A$ parameters can be confirmed in Figure~\ref{fig:Adelta_class}, where color represents the classification. 
Cores showing double peak profiles (dark orange and navy colors) have larger $|\delta_v|$ and $|A|$, while skew peak profiles (orange and blue) are found in cores showing smaller absolute values. Single profiles (green) are seen around 0 in both parameters. 
These results confirm that the quantitative asymmetry parameters and the morphological classification are in good agreement. 

\section{Discussion} \label{sec:discussion} 
\subsection{Infall Statistics and Cores Properties}
\begin{figure*}
    \centering
    \includegraphics[width=0.9\linewidth]{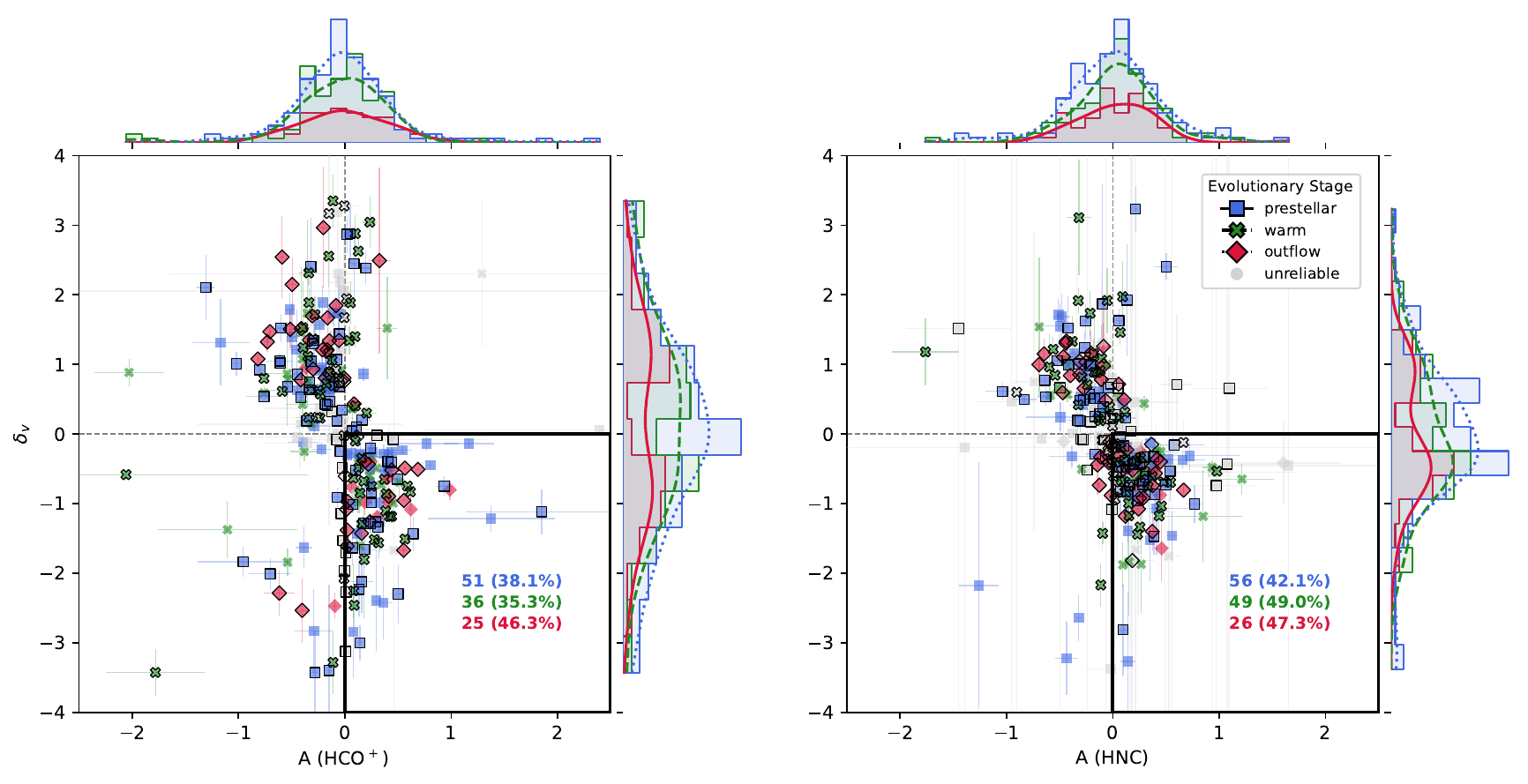}
    \caption{$\delta_v$ and $A$ scatter plots colored by cores' evolutionary stages. As with Figure~\ref{fig:A_delta_scatterplot}, a black edge means the core has a clear single peak in \dcop\ or N$_2$D$^+$. Cores with large uncertainties ($>$100\%) are colored gray.}
    \label{fig:Adelta_evol}
\end{figure*}

As they deviate from zero, both $A$ and $\delta_v$ indicate stronger inward/outward motion. 
Figure~\ref{fig:Adelta_evol} shows the $\delta_v$ and $A$ scatter plots colored by cores' evolutionary stages. 
The left and right panels show HCO$^+$ and HNC, respectively. 
The cores' evolutionary stages are classified in \citet{Morii24}. 
Cores associated with outflows (red diamonds) and with the detections of warm gas tracers (green crosses) are protostellar core candidates. 
Prestellar core candidates without any signs of protostars are plotted as blue squares. 

The difference among the evolutionary stages in the $A$ parameter lies in its dispersion, with prestellar cores showing greater variation. 
The median value does not differ significantly among the three evolutionary stages, but $A$ estimated from HNC peaks at $A>0$. 
Similarly, the $\delta_v$ distributions for all evolutionary stages take a peak $\delta_v<$-0.25 in \hnc, indicating the large population showing inward motion in \hnc. 
In $\delta_v$, a bimodal distribution is seen in protostellar cores, although prestellar cores show relatively smaller $|\delta_v|$ values with a single peak. 
The median value or the peak of $\delta_v$ distribution shifts to smaller values as evolution, implying that inward motion is more clearly seen in the protostellar phase. 
Since protostellar cores are more massive and denser \citep[][]{Morii26}, there is a weak correlation indicating that more massive and denser cores show relatively smaller $\delta_v$ and larger $A$, which is discussed in Appendix~\ref{sec:Adelta_phy}. 

\subsection{Line Profile Classification and Core Properties}\label{sec:class-phy}
\begin{figure*}
    \centering
    \includegraphics[width=0.8\linewidth]{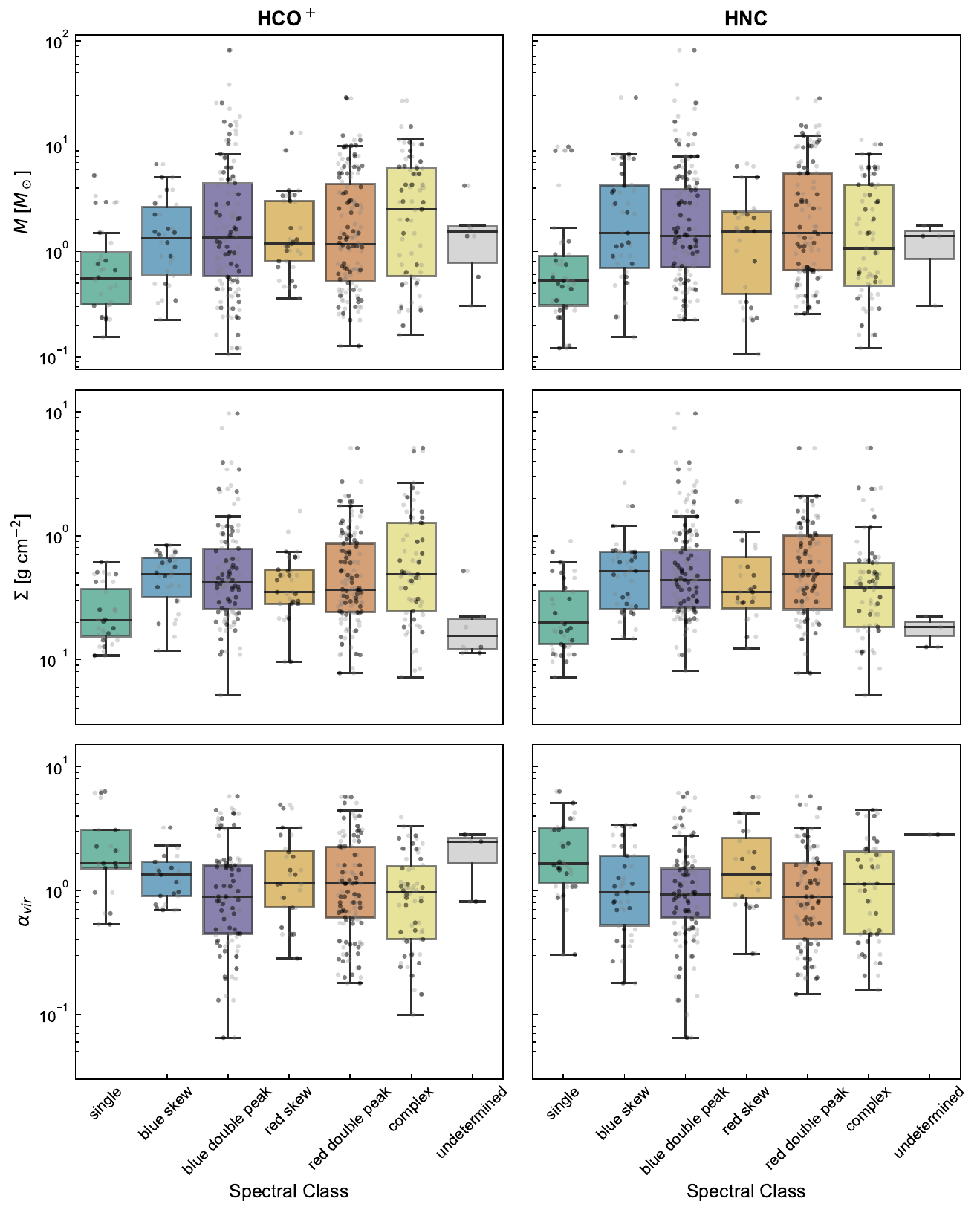} 
    \caption{Core physical parameters' variation against the spectral classification. The box plots show the 0th, 25th, 50th, 75th, and 100th percentiles, with outliers plotted individually. Cores with a single clear peak in \dcop\ or N$_2$D$^+$ are colored in black, while others are gray. }
    \label{fig:Core_phy_class_boxplot}
\end{figure*} 

Figure~\ref{fig:Core_phy_class_boxplot} shows the core physical properties distribution ($M$, $\Sigma$, $\alpha_{vir}$) along with the spectral classification. 
The left and right columns show the case of HCO$^+$ and HNC, respectively.  
Overall, the features seen in these plots are independent of the lines. 
Single and red skew profiles are seen in cores with low-mass, low-density, and relatively high virial parameters. 
Blue skew, blue double-peak, and red double-peak profiles are found in relatively more massive and dense cores with a small virial parameter ($\lesssim$2). 
Complex profiles are also found in low- to high-mass cores. 

\begin{figure*}
    \centering
    \includegraphics[width=0.8\textwidth]{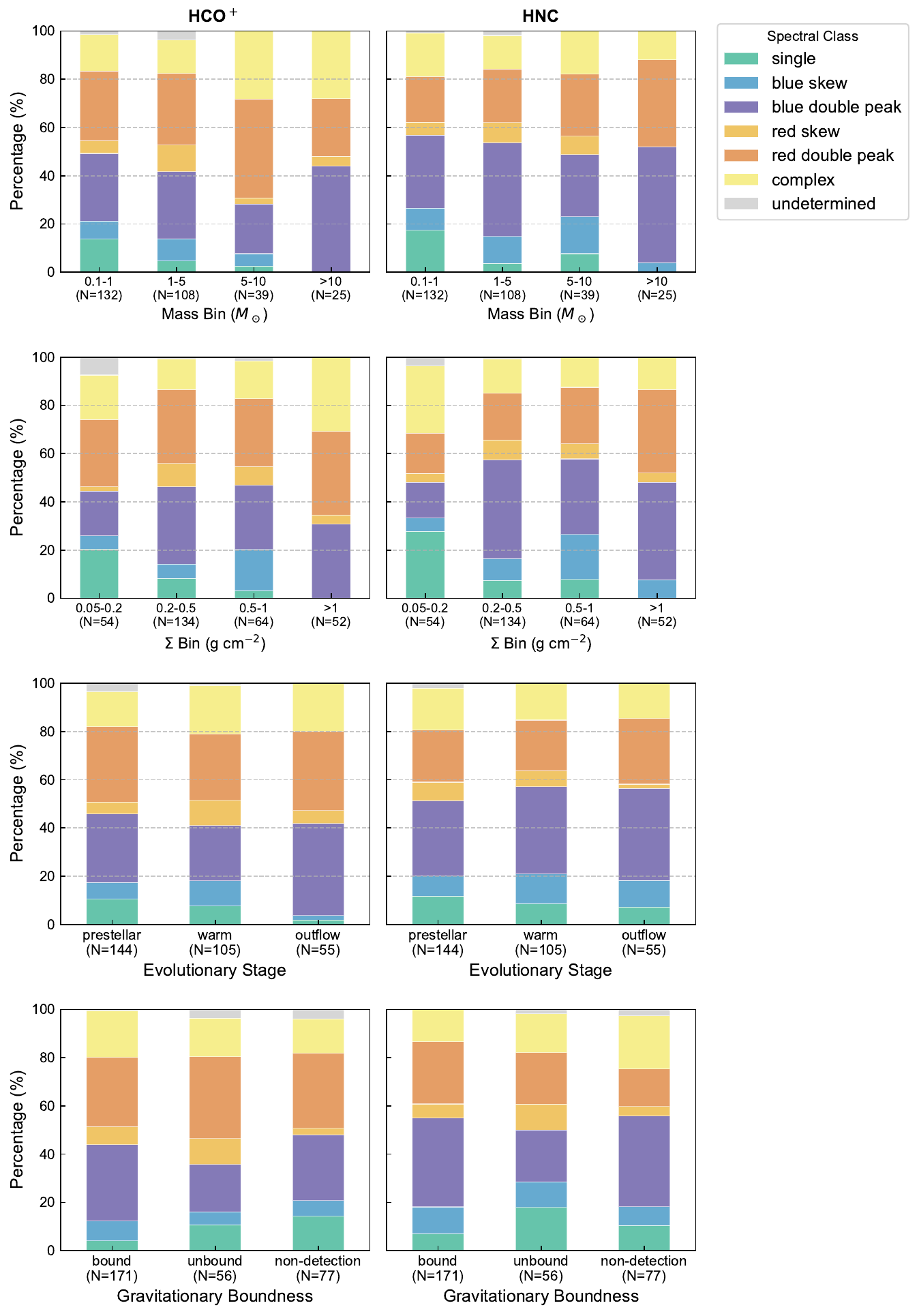} 
    \caption{The percentage of the spectral classes against the core mass, the surface density, the evolutionary stages, and gravitational boundness.}
    \label{fig:class_binplot}
\end{figure*}

Figure~\ref{fig:class_binplot} shows the percentage of each profile by dividing the sample into several mass and surface density bins in the first two rows.  
It clearly shows that the percentage of single-peak profiles decreases, while the percentage of blue double-peak profiles generally increases as mass and density increase. 
Considering \hcop\ is more affected by outflows and focusing on the \hnc\ plots, the blue asymmetry profile is the dominant category at the highest mass and density bin.  

It should be noted that, considering the physics behind the double peak profile, it is natural to see a clearer double peak profile in denser and more massive objects, where there is likely a steeper density and temperature gradient. Single or skew profiles in low-mass, low-density cores imply that those conditions may not be sufficient to make 3-2 lines optically thick (see also Section\ref{sec:line_detection}). 
However, as mentioned in the previous search for suitable lines tracing infall by numerical simulation \citep[e.g.,][]{Chira14, Jin-Jin21}, lower transitions would suffer more from the ambient gas around cores, raising the complexity to observe inward motion from low-mass, lower-density cores in massive clumps. 

The line profile classification and the relation with cores' evolutionary stage are summarized in the third row of Figure~\ref{fig:class_binplot}. 
Prestellar cores have the highest percentage of skew and single profiles. On the contrary, outflow-associated cores tend to show a double-peak profile (70\% when combining blue and red cases), indicating a broader line width. 
This implies that the blue asymmetry profile emerges from the prestellar stage and becomes increasingly dominant as evolution proceeds.  

Interestingly, we detected clear blue asymmetry profiles even in cores classified as gravitationally unbound (virial parameter $\alpha_{vir} > 2$). 
The bottom row of Figure~\ref{fig:class_binplot} shows the percentage for each category. 
Although the percentage ($\sim$20\%) is not as high as bound cores ($\sim$30-40\%), this is also confirmed in the scatter plot (right column of Figure~\ref{fig:A_delta_scatterplot}) where some unbound cores ($\alpha_{vir} \sim$2-4) show $A>0$ and $\delta<0$. 
This finding may challenge the simple interpretation of virial analysis, where external pressure is not considered. 
Or, as implied by \citet{Morii25}, this may be due to the overestimation of the non-thermal velocity dispersion caused by the broadening produced by local infall motions. 
It suggests that these cores may be undergoing collapse driven by external pressure (e.g., from the surrounding clump weight or feedback), rather than purely by self-gravity, as expected in accretion-driven star formation \citep[][]{Gomez21}, or turbulent fragmentation \citep{Ishihara25}. 

Our statistical analysis reveals that blue asymmetry is detected in nearly 60\% of the dense cores, excluding complex profiles. 
Numerical simulations by \citet{Smith12, Smith13} suggest that viewing angles, complex filamentary geometries, and large-scale gas flows can easily obscure infall signatures even in genuinely collapsing cores. 
The measured fraction is comparable to that reported by \citet{Chira14} for collapsing objects in simulations without feedback, implying that core-scale collapse is prevalent in 70 \textmu m-dark clumps. 
The fact that blue asymmetry is not ubiquitous likely reflects the combined effects of unfavorable viewing geometry and outflow contamination, rather than an absence of infall. 

\subsection{Hierarchical Inward Motion}
\begin{figure}
    \centering
    \includegraphics[width=0.8\linewidth]{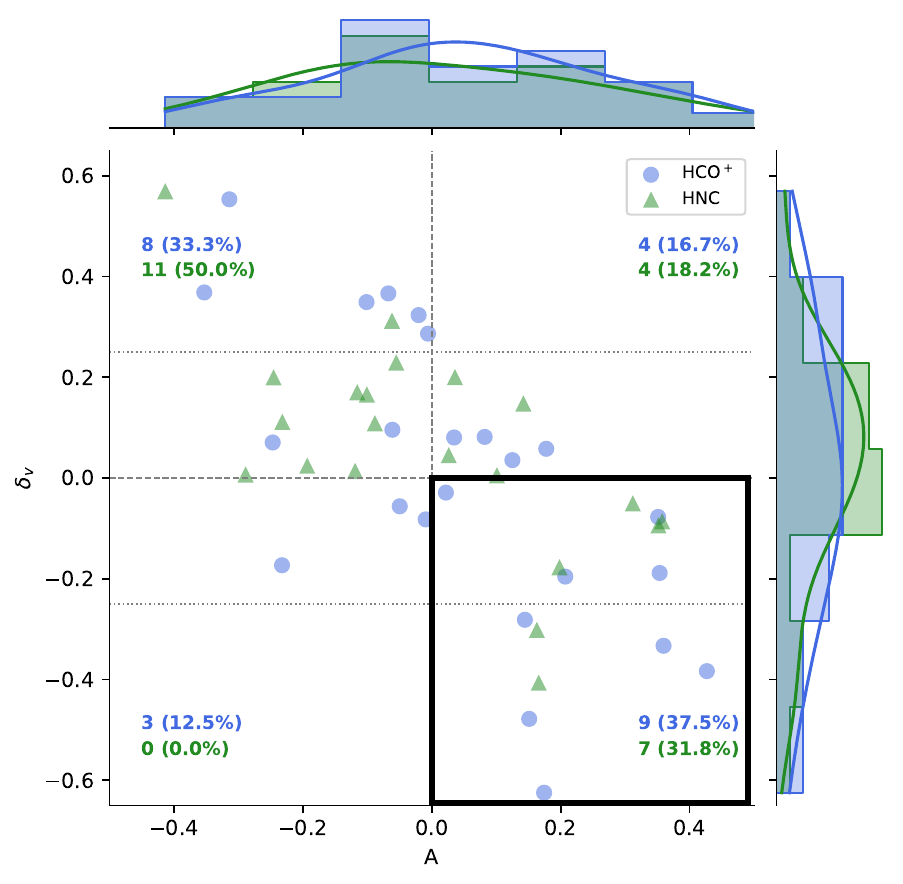}
    \caption{$\delta_v$ and $A$ scatter plots derived by HCO$^+$ (blue) and HNC line (green) for each clump. Figure style is the same with Figure~\ref{fig:A_delta_scatterplot}.}
    \label{fig:delta_A_cl}
\end{figure}

Gravitational collapse occurs not only at the core-scale but also at the clump-scale.
Using the total-power data, we measured $\delta_v$ and $A$ parameters at the clump-scale. 
The measured parameters are listed in Table~\ref{tab:infall-clump}. 
Figure~\ref{fig:delta_A_cl} is the same plot as Figure~\ref{fig:A_delta_scatterplot}, but for clump-scale profiles. 
The overall trend, such as the correlation between two parameters, is similar; however, the parameter space is much smaller. 
For example $|\delta_v|<0.6$ for clumps, but $\lesssim$2 for cores.
Compared to core-scale, clump-scale HCO$^+$ and HNC profiles are more likely to show a single peak, especially for HNC, resulting in small $\delta_v$ and $A$ variation. 
In clump-scale, HCO$^+$ rather shows blue-skew or double-peak profiles. 
This is consistent with \citet{Jackson26}, who report that HNC and HCN tend to show smaller asymmetries than \hcop\ due to their lower optical depth. 
Even at core scales, HNC becomes optically thick closer to the core center than \hcop\ \citep{Chira14}, so based on optical depth alone, \hcop\ would be expected to show a higher blue asymmetry fraction. 
The fact that the opposite trend is observed, though not significantly, is consistent with the earlier finding that \hcop\ profiles are more affected by outflows and diffuse gas flows, which can suppress or distort infall signatures. 

\begin{deluxetable}{lccccccc}
\tablecaption{Summary of Clump Properties\label{tab:infall-clump}}
\tablehead{
 \colhead{Region} & \colhead{n(H$_2$)\tablenotemark{a}} & \colhead{$f_{\rm proto}$\tablenotemark{b}} & \colhead{$N_{\rm core}$\tablenotemark{c}} & \multicolumn{2}{c}{$f_{\rm blue, all}$} & \colhead{$A_{\rm cl}$} & \colhead{$\delta_{\rm cl}$}  \\
 \colhead{} & \colhead{10$^4$ cm$^{-3}$} & \colhead{} & \colhead{} & \colhead{HCO$^+$} & \colhead{HNC} & \colhead{} & \colhead{}
}
\startdata
G016.97 & 4.19 & 0.38 & 12 & 0.33 & 0.33 & 0.14 & -0.28 \\
G018.80 & 4.09 & 0.75 & 6 & 0.17 & 0.17 & -0.07 & 0.37 \\
G018.93 & 2.04 & 0.47 & 5 & 0.20 & 0.20 & -0.31 & 0.55 \\
G022.69 & 1.75 & 0.42 & 9 & 0.11 & 0.11 & -0.25 & 0.07 \\
G023.47 & 8.39 & 0.53 & 12 & 0.75 & 0.75 & -0.00 & -0.11 \\
G024.01 & 11.52 & 0.62 & 13 & 0.46 & 0.46 & 0.17 & -0.62 \\
G024.52 & 2.91 & 0.65 & 18 & 0.33 & 0.33 & 0.43 & -0.38 \\
G025.16 & 3.74 & 0.56 & 13 & 0.46 & 0.46 & 0.15 & -0.48 \\
G028.54 & 2.86 & 0.39 & 16 & 0.25 & 0.25 & 0.12 & 0.04 \\
G028.56 & 8.36 & 0.71 & 26 & 0.23 & 0.23 & 0.03 & -0.03 \\
G028.92 & 2.44 & 0.56 & 7 & 0.57 & 0.57 & 0.15 & 0.06 \\
G030.70 & 1.73 & 0.52 & 10 & 0.60 & 0.60 & 0.03 & 0.08 \\
G030.91 & 6.17 & 0.42 & 10 & 0.50 & 0.50 & 0.08 & 0.08 \\
G033.33 & 1.21 & 0.33 & 7 & 0.29 & 0.29 & -0.35 & 0.37 \\
G034.13 & 1.58 & 0.13 & 10 & 0.00 & 0.00 & 0.36 & -0.08 \\
G034.73 & 4.48 & 0.58 & 10 & 0.00 & 0.00 & -0.01 & 0.29 \\
G036.66 & 5.04 & 0.54 & 9 & 0.44 & 0.44 & 0.21 & -0.20 \\
G305.79 & 9.30 & 0.47 & 27 & 0.41 & 0.41 & -0.06 & 0.10 \\
G327.11 & 4.10 & 0.19 & 11 & 0.09 & 0.09 & -0.09 & 0.35 \\
G332.96 & 1.94 & 0.10 & 7 & 0.14 & 0.14 & -0.07 & 0.32 \\
G333.48 & 4.04 & 0.20 & 18 & 0.67 & 0.67 & -0.05 & -0.06 \\
G333.52 & 10.93 & 0.71 & 28 & 0.43 & 0.43 & 0.36 & -0.33 \\
G337.54 & 5.64 & 0.53 & 11 & 0.18 & 0.18 & -0.23 & -0.17 \\
G340.39 & 2.90 & 0.10 & 9 & 0.44 & 0.44 & 0.35 & -0.19 \\
\enddata
\tablenotetext{a}{Estimated in \citet{Morii23}.}
\tablenotetext{b}{Fraction of the protostellar cores estimated in \citet{Morii24}.}
\tablenotetext{b}{Number of cores in the field-of-views of GLASES observations. }
\end{deluxetable}

\begin{figure*}
    \centering
    \includegraphics[width=0.99\linewidth]{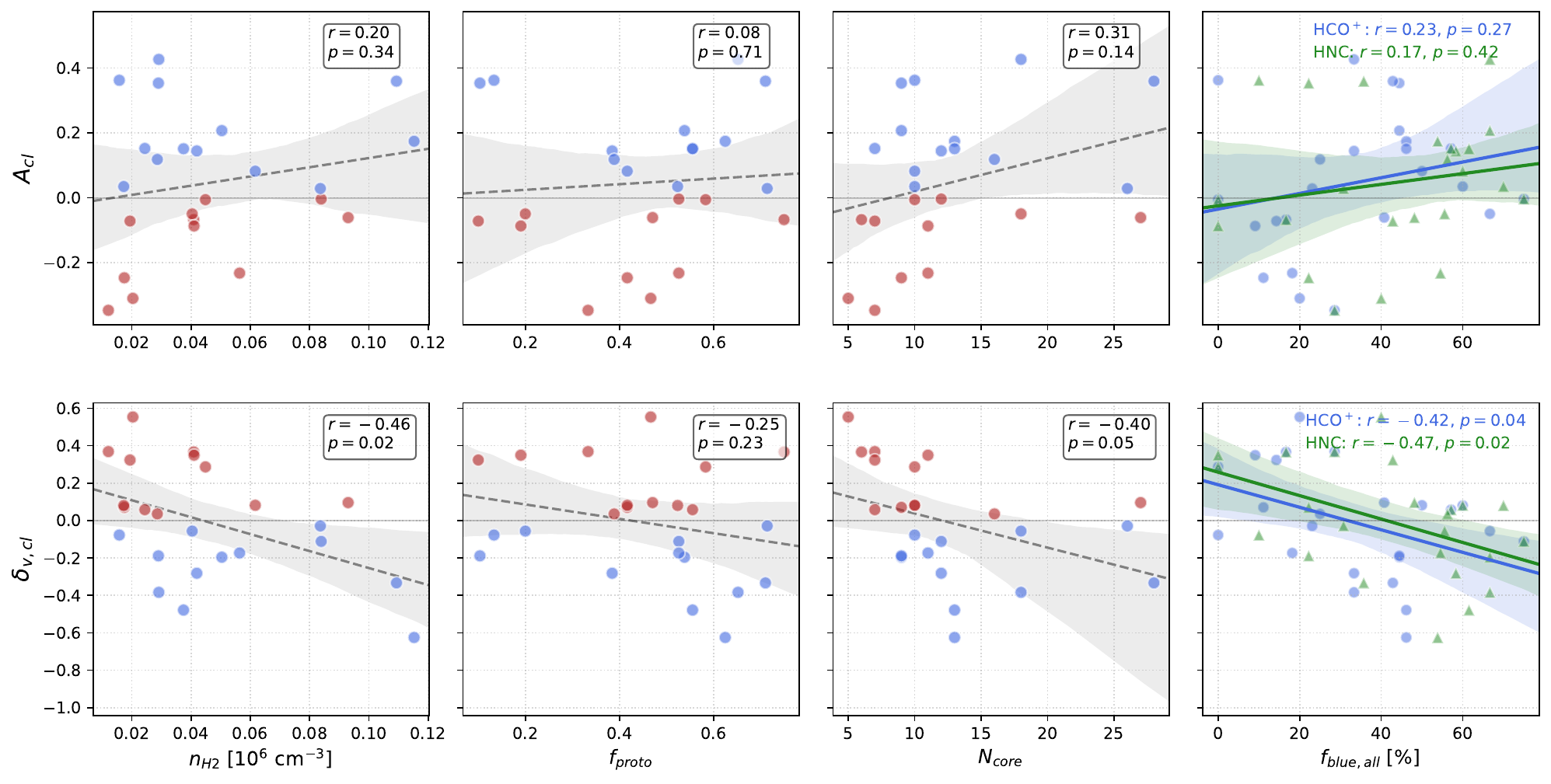}
    \caption{Scatter plots of the clump-scale $A_{\rm cl}$ and $\delta_{v, \rm{cl}}$ derived from \hcop\ line against the clump volume density ($n_{H_2}$), the protostellar core fraction, number of cores, and the fraction of cores showing blue asymmetric profiles ($f_{blue, all}$). Colors in the left three panels indicate whether the estimated $A_{\rm cl}$ or $\delta_{v, \rm{cl}}$ imply blue or red asymmetry profiles. The color in the right panel represents the species used for estimating $f_{blue, all}$. Regression plots are overlaid. Pearson correlation coefficients are denoted on the top-right. }
    \label{fig:delta_A_phy_cl}
\end{figure*}
We investigate the preferred physical conditions for detecting blue asymmetry profiles at the clump scale and their correlation with core-scale inward motion. 
We find some moderate correlation with clump density, protostellar core fraction and number of cores, which roughly traces the evolution of clumps. 
The left three panels of Figure~\ref{fig:delta_A_phy_cl} are scatter plots of $A$ and $\delta$ derived from HCO$^+$ with clump density, protostellar core fraction, and the number of cores. 
Data implying inward/outward motion are colored by blue and red, respectively. 
There is a moderate correlation, implying the clearer inward motion is seen in denser, likely more evolved clumps, with a larger number of cores. 

The two right panels of Figure~\ref{fig:delta_A_phy_cl} show the correlation found between clump-scale $A$ and $\delta$ derived from HCO$^+$ and core-scale profile classification. 
$f_{blue, all}$ is the fraction of blue-skew or blue-double peak profiles against the total number of cores in the region. 
All values used for the plots are in Table~\ref{tab:infall-clump}. 
The correlation between clump-scale and core-scale inward motion implies multiscale, hierarchical collapse \citep[e.g.,][]{Smith09, VazquezSemadeni24}: if clumps show inward motion, a higher percentage of cores show inward motion. 
However, we cannot fully rule out the possibility that the TP data partially reflect the cumulative emission from multiple cores rather than purely tracing the diffuse clump gas. Lower-transition observations would help to disentangle these contributions.

\subsection{Implications for Infall Velocity}
Before applying an infall model, we obtain a rough idea of the infall velocity and its variation with the peak difference or the peak velocity offset from the dip velocity. 
\begin{figure*}
    \centering
    \includegraphics[width=0.7\linewidth]{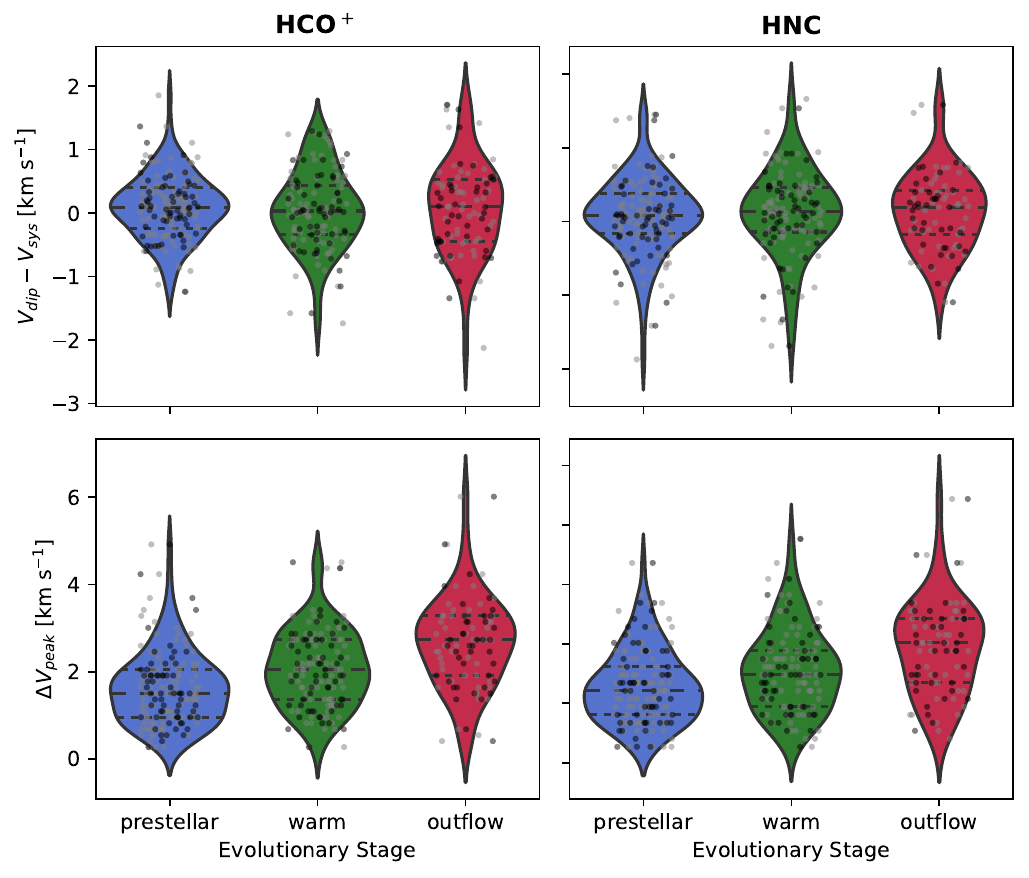}
    \caption{Distributions of dip velocity offset and peak separation across evolutionary stages. Top panels show the velocity difference between the absorption-like dip and the systemic velocity ($V_{\text{dip}} - V_{\text{sys}}$), serving as a proxy for infall velocity. Bottom panels show the velocity separation between the two peaks ($\Delta V_{\text{peak}}$), reflecting the line width and internal turbulence. The samples are categorized into three evolutionary stages: prestellar (blue), warm (green), and outflow-associated (red) cores. Individual core data points are overlaid on each violin plot.}
    \label{fig:Vdip_Vpeakdif}
\end{figure*}
Figure~\ref{fig:Vdip_Vpeakdif} shows the difference of $v_{\rm dip} - v_{\rm sys}$ and the velocity difference of the two peaks $\Delta V_{peak}$ if available, which roughly trace the infall velocity and the velocity dispersion, respectively. 
The top row indicates a weak trend of higher infall velocity with increasing evolution or mass. The median value is almost flat in both \hcop\ and \hnc\ plots. 
This implies a weak correlation between the evolutionary stage of cores and the expected infall velocity. 
The bottom row indicates an increase in velocity peak separation, implying a broader line width or stronger turbulence with evolution. 
However, as suggested by \citet{De_Vries05}, for example, more careful fitting with parameters of optical depth, temperature, and density, as well as infall velocity, is required to estimate the infall parameters. 

\section{Conclusions} \label{sec:conclusion}
We have presented the first large-scale statistical survey of gravitational infall signatures in high-mass dense cores using ALMA observations of optically thick molecular lines HCO$^+$ ($J$=3--2) and HNC ($J$=3--2). 
Our analysis of 304 dense cores within 24 massive 70~\textmu m-dark clumps from the GLASHES survey provides comprehensive evidence that core-scale collapse is prevalent in the earliest stages of high-mass star formation.
The main findings of this study are summarized as follows:

\begin{enumerate}
    \item 
    Blue-asymmetric line profiles, indicative of infalling gas motions, are detected in approximately 50--60\% of the cores using both the velocity difference parameter ($\delta_v < 0$) and the asymmetry parameter ($A > 0$). 
    More restrictive criteria ($\delta_v < -0.25$) yield detection rates of 40--45\%. 
    Spatially, these profiles are found not only inside cores but also around cores. 
    Considering that geometric effects and viewing angles can obscure infall signatures in genuinely collapsing cores, our results imply that infall motion is occurring around a substantial fraction of cores. 

    \item 
    The two optically thick tracers yield broadly consistent results, with HNC showing slightly higher detection rates of blue asymmetry (58\% vs.\ 50\% for $\delta_v < 0$). 
    This difference may reflect the fact that HCO$^+$ is more susceptible to contamination from outflow emission, which can introduce red-shifted components. 
    Despite these minor discrepancies, the overall agreement between the two tracers strengthens the robustness of our infall classifications.
    
    \item
    Approximately 60\% of the sample exhibits double-peaked profiles that can be described by two-component Gaussian fits, while 15\% show blue- or red-skewed profiles, 10\% show single peaks, and the remaining $\sim$15\% display complex morphologies. 
    Blue-asymmetric profiles become increasingly prevalent in more massive and denser cores, especially in cores with masses $M \gtrsim 10~M_{\odot}$ and surface densities $\Sigma \gtrsim 0.1$~g~cm$^{-2}$. 

    \item 
    Blue asymmetry is detected across all evolutionary stages, from prestellar cores to protostellar cores associated with outflows and warm gas emission lines. 
    As cores evolve, the profiles transition toward more pronounced double-peaked morphologies with larger velocity dispersions, suggesting increasing turbulence, feedback effects, or infall velocity. 
    The shift in the median $\delta_v$ toward more negative values with evolutionary stage indicates that infall signatures become more prominent in the protostellar phase.
    
    \item 
    Remarkably, blue-asymmetric profiles are detected in $\sim$20\% of cores with virial parameters $\alpha_{\rm vir} > 2$, which are nominally gravitationally unbound. 
    This finding implies that external compression from the surrounding core material may trigger collapse, or non-thermal velocity dispersion is overestimated due to the contribution of infall motion. 
    
    \item
    Analysis of clump-scale profiles reveals that inward motions occur simultaneously at both clump and core scales. 
    We find a moderate positive correlation between clump-scale infall signatures and the fraction of cores showing blue asymmetry within those clumps. 
    Clumps with higher densities, larger numbers of cores, and higher protostellar core fractions exhibit stronger infall signatures at both scales. 
    This hierarchical pattern supports the clump-fed accretion scenario. 

    \item 
    Preliminary estimates based on dip velocities and peak separations imply 
    a weak trend of increasing infall velocity and velocity dispersion with evolutionary stage and core mass. 
\end{enumerate}

In summary, our statistical survey demonstrates that gravitational collapse at the core scale is a prevalent and continuous process in high-mass star-forming regions. 
The detection of blue asymmetry across evolutionary stages, the infall signatures in nominally unbound cores, and the evidence for hierarchical inward motions from clump to core scales all underscore the dynamic and multi-scale nature of high-mass star formation.

\begin{acknowledgments}
We thank the anonymous referee for their careful reading and constructive comments, which improved the manuscript. 
PS was partially supported by a Grant-in-Aid for Scientific Research (KAKENHI Number JP23H01221) of JSPS. 
This paper makes use of the following ALMA data: 
ADS/JAO.ALMA\#2018.1.00299.S (PI: Y. Contreras), ADS/JAO.ALMA\#2023.1.01150.S (PI: K. Morii) and ADS/JAO.ALMA\#2024.1.01505.S (PI: K. Morii). 
ALMA is a partnership of ESO (representing its member states), NSF (USA) and NINS (Japan), together with NRC (Canada), NSTC and ASIAA (Taiwan), and KASI (Republic of Korea), in cooperation with the Republic of Chile. The Joint ALMA Observatory is operated by
ESO, AUI/NRAO and NAOJ.
The National Radio Astronomy Observatory is a facility of the National Science Foundation operated under cooperative agreement by Associated Universities, Inc.
\end{acknowledgments}

\facility {ALMA} 
\software{Astropy \citep{Astropy_13, Astropy18, Astropy22}, CASA \citep[][]{CASA22},  Matplotlib \citep{Hunter_matplotlib}, Numpy \citep{Harris20_numpy}, seaborn \citep{Waskom2021_seaborn}.} 

\appendix 
\section{Systemic Velocity Determination}\label{sec:Vsys}
To determine the systemic velocities ($V_{\rm thin}$) of the 304 cores in our sample, we first utilized the optically thin \dcop\ and \ntdp\ ($J$=3--2) lines. 
\citet{Morii24} applied a single Gaussian fit to the core-averaged spectra and estimated the properties for 227 cores. 
We excluded 5 cores from this subset due to marginal 3$\sigma$ detections and ambiguous systemic velocities, leaving 222 cores with robust \dcop or \ntdp measurements.

By carefully checking the spectra of these 222 cores, along with \dcn ($J$=3--2), we found that 39 cores host multiple velocity peaks. 
In particular, clumps G025.16, G028.56, G305.79, G333.48, G333.52, and G340.39 contain more than three cores showing multiple components. 
For these 39 cores, we determined the representative velocity from the brightest \dcop peak and added a flag to distinguish them from those with a single velocity component.

For the remaining 82 cores (i.e., 304 total $-$ 222) where neither \dcop\ nor \ntdp\ was securely detected, we fit an isolated hyperfine component of N$_2$H$^+$ ($J$=1--0, $F_1, F$ = 0,1--1,2 $\nu$= 93.176265 GHz). Through this method, we obtained $V_{\rm thin}$ for 68 of the 82 cores. Consequently, we successfully derived velocity information from optically thin lines for a total of 290 cores (222 from \dcop\ or \ntdp\ + 68 from N$_2$H$^+$). 
For the final 14 cores without any thin line detections, we assigned the average $V_{\rm thin}$ of their respective regions.

To reduce the impact of ambiguous $V_{\rm thin}$ measurements in our analysis, we flagged a total of 121 cores: the 82 cores lacking \dcop\ or \ntdp\ detections and the 39 cores exhibiting multiple components. This leaves a highly reliable sub-sample of 183 cores. The 121 flagged cores are highlighted in most of the figures. 
The derived $V_{\rm thin}$ for each core is summarized in Table~\ref{tab:infall}.

\section{Total--power Data Combination Effects}\label{sec:TP-combination}
\begin{figure}
    \centering
    \includegraphics[width=0.99\linewidth]{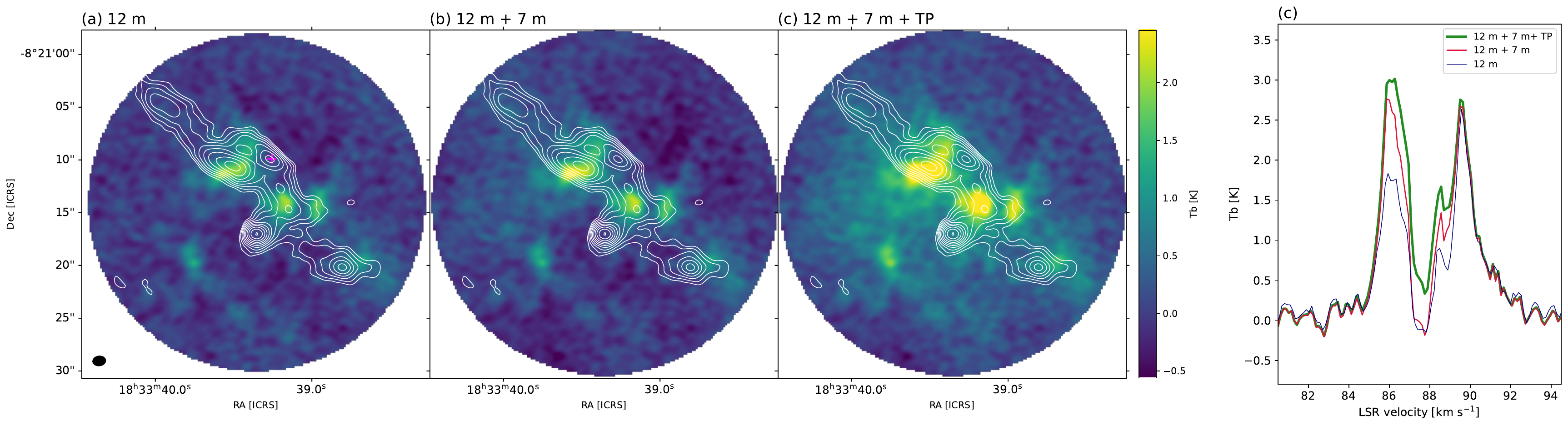}
    \caption{Comparison of the spatial distribution of the emission and the line spectra with and without Total Power data. The left three images are one slice of the cube at the same channel and color level, but (a) 12 m-only, (b) 12 m + 7 m, and (c) 12 m + 7 m + TP, respectively. The line spectra on panel (d) show the direct comparison of the spectra at the pixel where the magenta cross is in panel (a). Green, red, and navy lines represent the different cases of the data combination. White contours are the same with Figure~\ref{fig:mom0_cont}.} 
    \label{fig:TP-combination}
\end{figure}

In our data analysis, we combine total power data to recover the extended emission lost due to the interferometer. 
To illustrate the significance of this process, in this Appendix we compare the data before and after the TP combination. 
Figure~\ref{fig:TP-combination} shows the difference in the spatial distribution and profile of the emission. 
From left to right, the images show a single channel of a 12-m data cube, a 12-m + 7-m combined data cube, and a 12-m + 7-m + TP combined data cube, respectively. 
The color scale of the three images is the same. Clearly, the peak intensity and spatial distribution differ. Panel (a) shows negative components around the continuum cores, while panel (c) shows more extended emission than the continuum. 
The spectrum in panel (d) also clearly illustrates the differences. There is no significant difference far from the line center ($v_{\rm lsr} \sim$88 km s$^{-1}$), however, differences in intensity of a factor of a few exist around $v_{lsr} \sim$85.5–90 km\,s$^{-1}$. 
Regardless of the infall profile classification or modeling based on the line profile, it is clear that the combination of the 7 m and TP matters and can alter the conclusion. 

\section{Line Profile Classification} \label{sec:profile-classification}
The spectral profile was classified using a combination of peak-based criteria, dip significance, and Gaussian decomposition parameters.  
First, all local intensity maxima with S/N $>$3 were identified within a velocity range of $\pm$ 4$\times \sigma_{\rm thin}$ around the systemic velocity ($V_{\text{sys}}$). 
A peak was considered significant only if its amplitude exceeded the rms noise level by a factor of 3. 
The significant peaks were then divided into blue-shifted and red-shifted groups. 
If no reliable peak velocity could be determined, the profile was labeled as \textit{undetermined}. 
When multiple significant peaks were present on either side, we selected only the strongest peak (i.e., the peak with the maximum intensity) as the representative blue or red component. 

For profiles with $N_{\text{peak}} = 2$ (i.e., both blue and red peaks exist), a double-Gaussian model was applied immediately. 
For other cases, a single-Gaussian fit was first performed. 
If the resulting residuals exceeded $5\sigma_{\text{rms}}$ noise level, the model was upgraded to a double-Gaussian to account for hidden sub-structures. 
When a two-component Gaussian fit was available (six fit parameters), for each fitted Gaussian, the observed peak velocities close to each centroid velocity were identified. 
The relative amplitudes at these matched peak velocities were then compared to determine whether the dominant component lies on the blue- or red-shifted side relative to the systemic velocity.   
This matching procedure is necessary because extended components, such as outflows, can produce broad Gaussian fits whose centroids are displaced from the local spectral maxima. 

Then, the velocity separation between the two Gaussian centroids ($\Delta v$) and the sum of their widths \(\sigma_{1}+\sigma_{2}\) were used to distinguish morphological types:
\begin{itemize}
  \item If \(\Delta v < 0.2\ \mathrm{km\ s^{-1}}\), the profile was classified as \textit{single}.
  \item If \(\Delta v > \sigma_{1}+\sigma_{2}\), the profile was classified as a \textit{double peak}; the dominant side (blue or red) was indicated according to the matched amplitudes.
  \item Otherwise, the profile was classified as \textit{blue-skewed} or \textit{red-skewed} depending on which side exhibited the stronger matched component.
  \item If either Gaussian width satisfied \(\sigma>2.0\ \mathrm{km\ s^{-1}}\), the profile was labeled as \textit{complex}.
\end{itemize}

When only a single-Gaussian fit was available, the profile was classified as \textit{single}.  
If Gaussian fitting was not available or not conclusive, classification was done based on the peak velocity information: two detected peaks with separation \(<0.2\ \mathrm{km\ s^{-1}}\) were labeled \textit{single}; peaks separated by \(>2.0\ \mathrm{km\ s^{-1}}\) were labeled \textit{complex}; intermediate separations were assigned \textit{blue-skewed} or \textit{red-skewed} according to the relative peak intensities.

\begin{figure}
    \centering
    \includegraphics[width=0.8\linewidth]{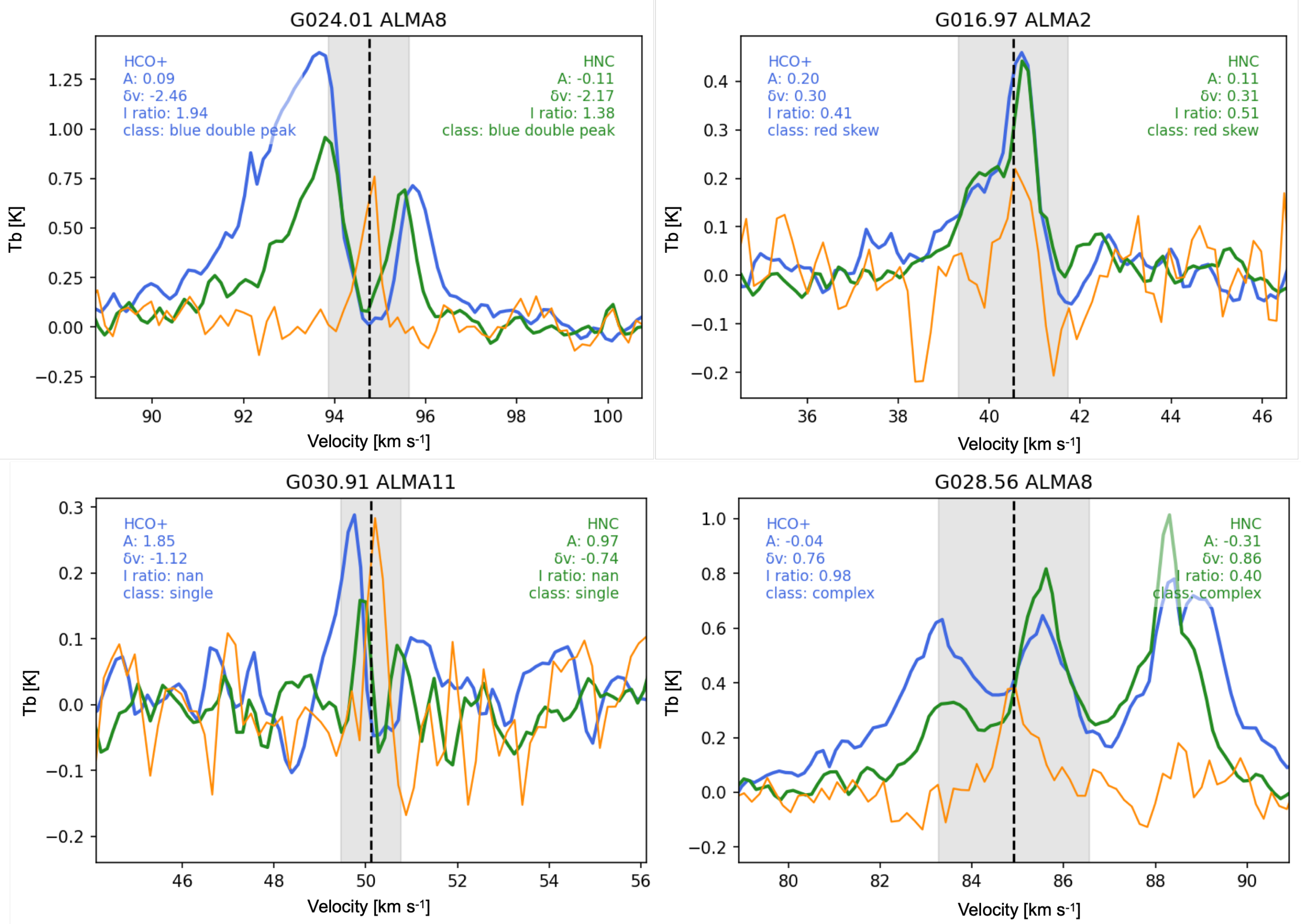}
    \caption{Examples of the profiles for four cores showing various classifications. Blue and green colors represent \hcop\ and \hnc, respectively. Orange line represents the optically thin tracer (i.e., \dcop\ or \ntdp, which is multiplied by 2 in intensity. $V_{\rm thin}$ is denoted as the black dashed line, and the gray shaded areas show the range where $A$ is calculated ($V_{\rm thin} - 2\Delta V_{\rm thin}$). } 
    \label{fig:class_profile}
\end{figure}

Figure~\ref{fig:class_profile} shows some examples of the profiles with classified classes, such as double peak, skew, single, and complex, respectively. 
As seen in the bottom left panel, \textit{single} may be due to the lack of sensitivity to detect the counterpart. Most \textit{complex} profiles are affected by the multi-velocity components implied by the optically thin tracer, but as shown in the bottom right, there is a core with a single peak in optically thin line but with multiple peaks in optically thick lines. 

\begin{figure}
    \centering
    \includegraphics[width=0.6\linewidth]{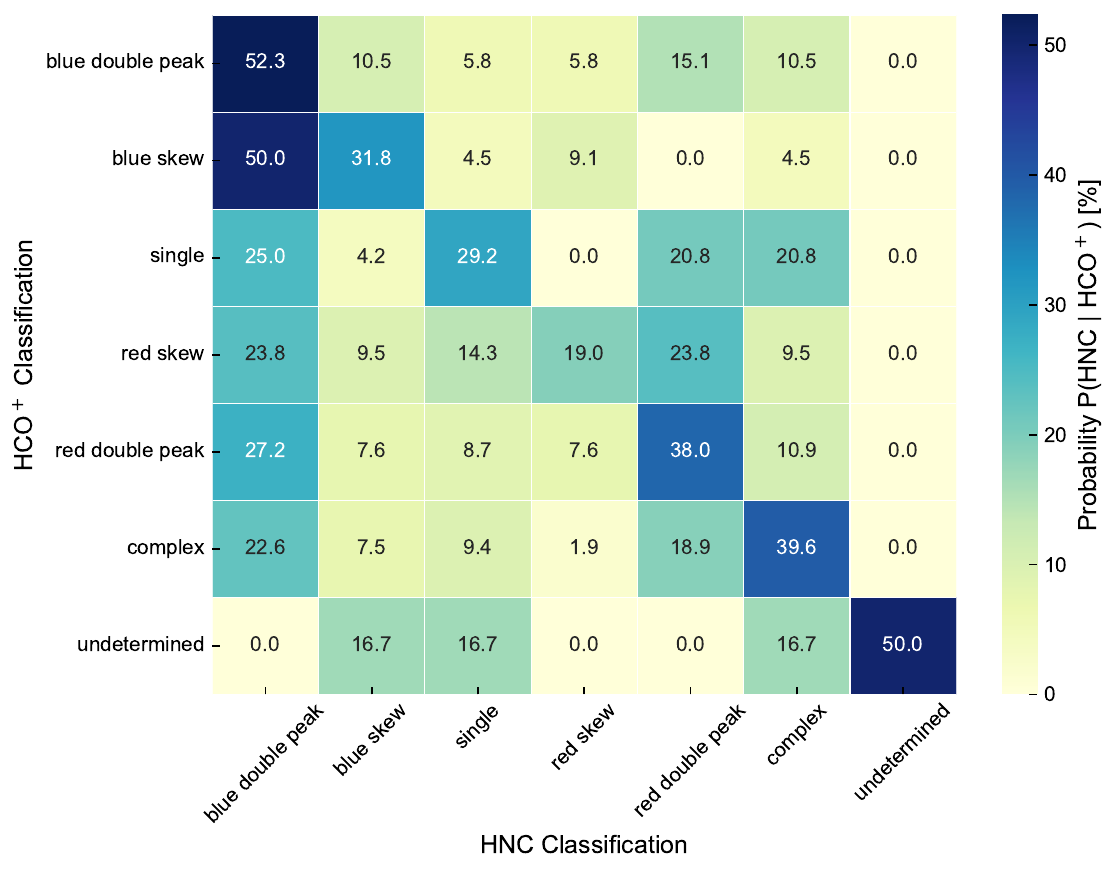}
    \caption{Correlation plot of line classification between HCO$^+$ and HNC.} 
    \label{fig:class_corr}
\end{figure}

Figure~\ref{fig:class_corr} shows the correlation plot of line classification between HCO$^+$ and HNC, indicating the good correlation between the two lines, but not always the same. 
In particular, $\sim$30\% of cores showing blue double profile in HNC show red double peak in HCO$^+$. 
This may reflect the contribution from outflow according to the higher chance of spectra with tail/wing structures seen in HCO$^+$. 
Some showing a single peak profile in HNC, which tend to be low-mass or less dense cores, show a skew profile in HCO$^+$, implying the initial signature of inward motion is easier to see in HCO$^+$  than HNC. 
More detailed discussion of these classifications and cores' physical properties is in Section~\ref{sec:class-phy}. 

\section{Optical Depth Estimation of \hcop\ ($J=3-2$)} \label{sec:opticaldepth}
To estimate the peak optical depth ($\tau$) of the \hcop\ ($3–2$) transition, we utilized the rarer isotopologue HC$^{18}$O$^+$ ($J=3-2$) as an optically thin reference. 
We analyzed three representative regions (G022.69, G023.47, and G030.91), covering a total of 30 cores that span all line-profile categories discussed in this work (prestellar, protostellar, double-peaked, blue/red-skewed, and single-peaked). 
Both data sets used the TP+12m+7m combined products. 
Assuming Local Thermodynamic Equilibrium (LTE)
, the total column density is given by:
\begin{equation}
    N = \frac{8\pi \nu^3}{c^3} \frac{ Q(T_{\rm ex})}{g_{\rm u} A_{\rm ul}}
    \frac{ {\rm exp} (E_{\rm u}/T_{\rm ex}) }{[{\rm exp} (h\nu/k_B T_{\rm ex})-1]} \int \tau dv
    \approx \frac{8\pi \nu^3}{c^3} \frac{ Q(T_{\rm ex})}{g_{\rm u} A_{\rm ul}}
    \frac{ {\rm exp} (E_{\rm u}/T_{\rm ex})}{[{\rm exp} (h\nu/k_B T_{\rm ex})-1]} \frac{W}{J(T_{\rm ex}) - J(T_{\rm bg})}
\end{equation}
where the approximation in the second line holds under the optically thin assumption ($\tau \ll1$) and $J(T) = \frac{h\nu/k_B}{{\rm exp}(h\nu/k_B T)-1}$. 
Here $W= T_{\rm b,peak}\,\Delta v_{\rm FWHM}\sqrt{\frac{\pi}{4\ln 2}}$ is the integrated intensity estimated from a Gaussian fit to the HC$^{18}$O$^+$ line, $T_{\rm b,peak}$ is the peak brightness temperature, $\Delta v_{\rm FWHM}$ is the full width at half maximum derived from the Gaussian fit, $A_{\rm ul}$ is the Einstein A coefficient, $g_{\rm u}$ is the upper-level degeneracy, $E_{\rm u}$ is the upper-level energy, $Q(T_{\rm ex})$ is the partition function, and $T_{\rm bg} = 2.73$ K is the cosmic microwave background temperature. 
The spectroscopic parameters adopted for HC$^{18}$O$^+$ ($J=3-2$) are $\nu =$255.488 GHz, $A_{\rm ul}=1.265\times10^{-3}$ s$^{-1}$, $E_{\rm u} = 24.52$ K, and $g_{\rm u} = 7$. 
We adopted the excitation temperature $T_{\rm ex} =$18.0, 13.9, and 12.4 K  for G022.69, G023.47, and G030.91, derived from Herschel SED fitting \citep{Morii23}, and assumed that both isotopologues share the same $T_{\rm ex}$. It should be noted that if the true excitation temperature of \hcop\ is lower than our assumed $T_{ex}$, the derived optical depth $\tau$(HCO$^+$) would be underestimated. 

The total \hcop\ column density was obtained by applying the Galactocentric distance-dependent isotopic abundance ratio:
\begin{equation}
    \frac{[{\rm ^{16}O}]}{[{\rm ^{18}O}]} = 58.8\,R_{\rm GC}\,{\rm (kpc)} + 37.1
\end{equation}
following \citet{Wilson94}. 
The adopted Galactocentric distances and resulting abundance ratios are $R_{\rm GC} = $4.38, 4.19, and 5.81 kpc for G022.69, G023.47, and G030.91 \citep{Whitaker17}, corresponding to ratios of $\sim$295, $\sim$284, and $\sim$379, respectively. 
The peak optical depth was then derived by inverting the standard LTE column density expression:
\begin{equation}
\tau_{\rm HCO^+} = N_{\rm HCO^+} / \left[\dfrac{8\pi\nu^3}{c^3} \cdot \dfrac{Q(T_{\rm ex})}{g_{\rm u} A_{\rm ul}} \cdot \dfrac{{\rm exp}(E_{\rm u}/T_{\rm ex)}}{{\rm exp}(h\nu/k_B T_{\rm ex})-1} \cdot \Delta v_{\rm FWHM}\sqrt{\dfrac{\pi}{4\ln 2}}\right]
\end{equation}
where the spectroscopic parameters for \hcop\ (3–2) are $\nu = 267.566$ GHz, $A_{\rm ul} = 1.438\times10^{-3}$ s$^{-1}$, $E_{\rm u} = 25.68$ K, $g_{\rm u} = 7$, and the partition function is approximated as $Q(T_{\rm ex}) = 0.4671\,T_{\rm ex} + 0.3445$.

Of the 30 cores analyzed, HC$^{18}$O$^+$ ($J=3-2$) was detected in 14 cores, yielding optical depths in the range of 1.7–11.4 (mean=6.3), confirming that \hcop ($3–2$) is optically thick across a wide range of core properties. 
The majority of these detections correspond to cores exhibiting double-peaked \hcop profiles. 
For the remaining 16 cores where HC$^{18}$O$^+$ was not detected, we used the $3\sigma$ rms noise level of the spectra to derive upper limits, obtaining $\tau_{\rm HCO^+} < 0.8 –13.7$ (mean upper limit = 6.7). 
None of the non-detections indicate clearly optically thin conditions, and optical thickness cannot be excluded in any of these cases. 

\section{Infall Parameters and Physical Parameters} \label{sec:Adelta_phy}
\begin{figure*}
    \centering
    \includegraphics[width=0.8\linewidth]{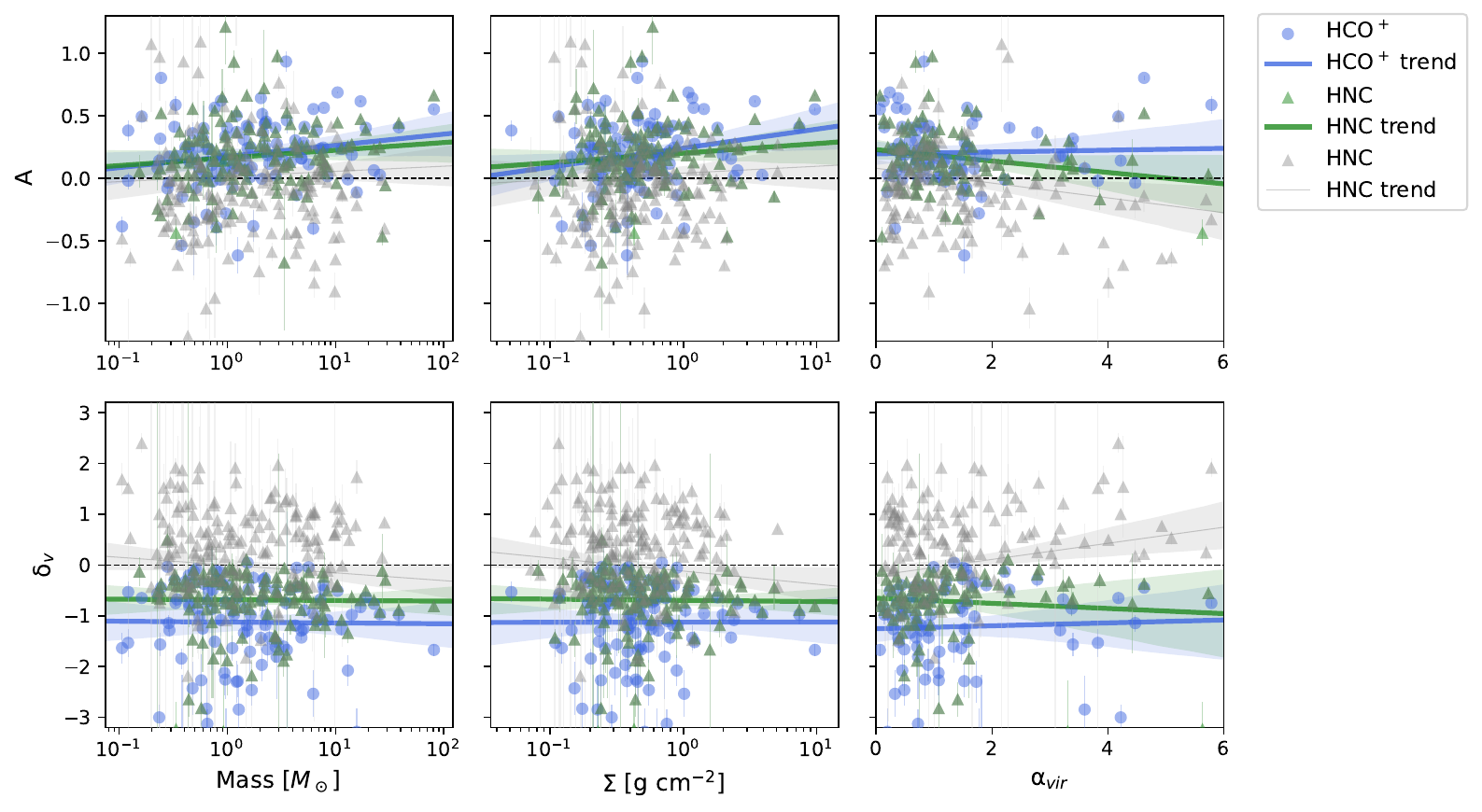}
    \caption{$\delta_v$ and $A$ against core mass, surface density, and virial parameter. Colors are the same as Figure~\ref{fig:A_delta_scatterplot}. Lines are regression plots. Cores with blue asymmetry line profiles (i.e., a blue double peak or blue skew) are shown in color. Gray triangles and regression plots show the parameters of all cores measured by the HNC line. }
    \label{fig:Adelta_MS}
\end{figure*}

We investigated if there is any correlation between these parameters and cores physical properties. 
Figure~\ref{fig:Adelta_MS} shows the scatter plots of $A$ and $\delta_v$ against core mass (left), surface density (middle), and virial parameters (right). 
Cores with blue asymmetry line profiles (i.e., a blue double peak or blue skew) are shown in color, where color represents the molecule, same as Figure~\ref{fig:A_delta_scatterplot}.   
The lines overlaid are regression plots. 
There is a trend of A-increase as core mass and surface density increase, and a decrease in the virial parameter. 
The $\delta_v$ parameter does not significantly change with core mass, density, and virial parameter. 
However, there are some trends when all cores are considered (gray triangles for the HNC case). 
These tendencies imply that more inward motions are seen from more massive, denser, and bound cores, which is consistent with the main conclusion from the main text. 

\bibliography{references}

@ARTICLE{Mardones97,
       author = {{Mardones}, D. and {Myers}, P.~C. and {Tafalla}, M. and {Wilner}, D.~J. and {Bachiller}, R. and {Garay}, G.},
        title = "{A Search for Infall Motions toward Nearby Young Stellar Objects}",
      journal = {\apj},
         year = 1997,
        month = nov,
       volume = {489},
       number = {2},
        pages = {719-733},
          doi = {10.1086/304812},
archivePrefix = {arXiv},
       eprint = {astro-ph/9707011},
 primaryClass = {astro-ph},
       adsurl = {https://ui.adsabs.harvard.edu/abs/1997ApJ...489..719M}
}

@ARTICLE{Myers96,
       author = {{Myers}, P.~C. and {Mardones}, D. and {Tafalla}, M. and {Williams}, J.~P. and {Wilner}, D.~J.},
        title = "{A Simple Model of Spectral-Line Profiles from Contracting Clouds}",
      journal = {\apjl},
         year = 1996,
        month = jul,
       volume = {465},
        pages = {L133},
          doi = {10.1086/310146},
       adsurl = {https://ui.adsabs.harvard.edu/abs/1996ApJ...465L.133M}
}

@ARTICLE{Zhou93,
       author = {{Zhou}, Shudong and {Evans}, II, Neal J. and {Koempe}, Carsten and {Walmsley}, C.~M.},
        title = "{Evidence for Protostellar Collapse in B335}",
      journal = {\apj},
         year = 1993,
        month = feb,
       volume = {404},
        pages = {232},
          doi = {10.1086/172271},
       adsurl = {https://ui.adsabs.harvard.edu/abs/1993ApJ...404..232Z}
}

@ARTICLE{Morii25,
       author = {{Morii}, Kaho and {Sanhueza}, Patricio and {Csengeri}, Timea and {Nakamura}, Fumitaka and {Bontemps}, Sylvain and {Garay}, Guido and {Zhang}, Qizhou},
        title = "{Global and Local Infall in the ASHES Sample (GLASHES). I. Pilot Study in G337.541}",
      journal = {\apj},
         year = 2025,
        month = feb,
       volume = {979},
       number = {2},
          eid = {233},
        pages = {233},
          doi = {10.3847/1538-4357/ada27f},
archivePrefix = {arXiv},
       eprint = {2412.17901},
 primaryClass = {astro-ph.GA},
       adsurl = {https://ui.adsabs.harvard.edu/abs/2025ApJ...979..233M}
}

@article{Waskom2021_seaborn,
    doi = {10.21105/joss.03021},
    url = {https://doi.org/10.21105/joss.03021},
    year = {2021},
    publisher = {The Open Journal},
    volume = {6},
    number = {60},
    pages = {3021},
    author = {Michael L. Waskom},
    title = {seaborn: statistical data visualization},
    journal = {Journal of Open Source Software}
 }

@ARTICLE{Hunter_matplotlib,
       author = {{Hunter}, John D.},
        title = "{Matplotlib: A 2D Graphics Environment}",
      journal = {Computing in Science and Engineering},
         year = 2007,
        month = may,
       volume = {9},
       number = {3},
        pages = {90-95},
          doi = {10.1109/MCSE.2007.55},
       adsurl = {https://ui.adsabs.harvard.edu/abs/2007CSE.....9...90H}
}

@ARTICLE{Astropy22,
       author = {{Astropy Collaboration} and {Price-Whelan}, Adrian M. and {Lim}, Pey Lian and {Earl}, Nicholas and {Starkman}, Nathaniel and {Bradley}, Larry and {Shupe}, David L. and {Patil}, Aarya A. and {Corrales}, Lia and {Brasseur}, C.~E. and {N{\"o}the}, Maximilian and {Donath}, Axel and {Tollerud}, Erik and {Morris}, Brett M. and {Ginsburg}, Adam and {Vaher}, Eero and {Weaver}, Benjamin A. and {Tocknell}, James and {Jamieson}, William and {van Kerkwijk}, Marten H. and {Robitaille}, Thomas P. and {Merry}, Bruce and {Bachetti}, Matteo and {G{\"u}nther}, H. Moritz and {Aldcroft}, Thomas L. and {Alvarado-Montes}, Jaime A. and {Archibald}, Anne M. and {B{\'o}di}, Attila and {Bapat}, Shreyas and {Barentsen}, Geert and {Baz{\'a}n}, Juanjo and {Biswas}, Manish and {Boquien}, M{\'e}d{\'e}ric and {Burke}, D.~J. and {Cara}, Daria and {Cara}, Mihai and {Conroy}, Kyle E. and {Conseil}, Simon and {Craig}, Matthew W. and {Cross}, Robert M. and {Cruz}, Kelle L. and {D'Eugenio}, Francesco and {Dencheva}, Nadia and {Devillepoix}, Hadrien A.~R. and {Dietrich}, J{\"o}rg P. and {Eigenbrot}, Arthur Davis and {Erben}, Thomas and {Ferreira}, Leonardo and {Foreman-Mackey}, Daniel and {Fox}, Ryan and {Freij}, Nabil and {Garg}, Suyog and {Geda}, Robel and {Glattly}, Lauren and {Gondhalekar}, Yash and {Gordon}, Karl D. and {Grant}, David and {Greenfield}, Perry and {Groener}, Austen M. and {Guest}, Steve and {Gurovich}, Sebastian and {Handberg}, Rasmus and {Hart}, Akeem and {Hatfield-Dodds}, Zac and {Homeier}, Derek and {Hosseinzadeh}, Griffin and {Jenness}, Tim and {Jones}, Craig K. and {Joseph}, Prajwel and {Kalmbach}, J. Bryce and {Karamehmetoglu}, Emir and {Ka{\l}uszy{\'n}ski}, Miko{\l}aj and {Kelley}, Michael S.~P. and {Kern}, Nicholas and {Kerzendorf}, Wolfgang E. and {Koch}, Eric W. and {Kulumani}, Shankar and {Lee}, Antony and {Ly}, Chun and {Ma}, Zhiyuan and {MacBride}, Conor and {Maljaars}, Jakob M. and {Muna}, Demitri and {Murphy}, N.~A. and {Norman}, Henrik and {O'Steen}, Richard and {Oman}, Kyle A. and {Pacifici}, Camilla and {Pascual}, Sergio and {Pascual-Granado}, J. and {Patil}, Rohit R. and {Perren}, Gabriel I. and {Pickering}, Timothy E. and {Rastogi}, Tanuj and {Roulston}, Benjamin R. and {Ryan}, Daniel F. and {Rykoff}, Eli S. and {Sabater}, Jose and {Sakurikar}, Parikshit and {Salgado}, Jes{\'u}s and {Sanghi}, Aniket and {Saunders}, Nicholas and {Savchenko}, Volodymyr and {Schwardt}, Ludwig and {Seifert-Eckert}, Michael and {Shih}, Albert Y. and {Jain}, Anany Shrey and {Shukla}, Gyanendra and {Sick}, Jonathan and {Simpson}, Chris and {Singanamalla}, Sudheesh and {Singer}, Leo P. and {Singhal}, Jaladh and {Sinha}, Manodeep and {Sip{\H{o}}cz}, Brigitta M. and {Spitler}, Lee R. and {Stansby}, David and {Streicher}, Ole and {{\v{S}}umak}, Jani and {Swinbank}, John D. and {Taranu}, Dan S. and {Tewary}, Nikita and {Tremblay}, Grant R. and {de Val-Borro}, Miguel and {Van Kooten}, Samuel J. and {Vasovi{\'c}}, Zlatan and {Verma}, Shresth and {de Miranda Cardoso}, Jos{\'e} Vin{\'\i}cius and {Williams}, Peter K.~G. and {Wilson}, Tom J. and {Winkel}, Benjamin and {Wood-Vasey}, W.~M. and {Xue}, Rui and {Yoachim}, Peter and {Zhang}, Chen and {Zonca}, Andrea and {Astropy Project Contributors}},
        title = "{The Astropy Project: Sustaining and Growing a Community-oriented Open-source Project and the Latest Major Release (v5.0) of the Core Package}",
      journal = {\apj},
         year = 2022,
        month = aug,
       volume = {935},
       number = {2},
          eid = {167},
        pages = {167},
          doi = {10.3847/1538-4357/ac7c74},
archivePrefix = {arXiv},
       eprint = {2206.14220},
 primaryClass = {astro-ph.IM},
       adsurl = {https://ui.adsabs.harvard.edu/abs/2022ApJ...935..167A}
}

@ARTICLE{Astropy18,
       author = {{Astropy Collaboration} and {Price-Whelan}, A.~M. and {Sip{\H{o}}cz}, B.~M. and {G{\"u}nther}, H.~M. and {Lim}, P.~L. and {Crawford}, S.~M. and {Conseil}, S. and {Shupe}, D.~L. and {Craig}, M.~W. and {Dencheva}, N. and {Ginsburg}, A. and {VanderPlas}, J.~T. and {Bradley}, L.~D. and {P{\'e}rez-Su{\'a}rez}, D. and {de Val-Borro}, M. and {Aldcroft}, T.~L. and {Cruz}, K.~L. and {Robitaille}, T.~P. and {Tollerud}, E.~J. and {Ardelean}, C. and {Babej}, T. and {Bach}, Y.~P. and {Bachetti}, M. and {Bakanov}, A.~V. and {Bamford}, S.~P. and {Barentsen}, G. and {Barmby}, P. and {Baumbach}, A. and {Berry}, K.~L. and {Biscani}, F. and {Boquien}, M. and {Bostroem}, K.~A. and {Bouma}, L.~G. and {Brammer}, G.~B. and {Bray}, E.~M. and {Breytenbach}, H. and {Buddelmeijer}, H. and {Burke}, D.~J. and {Calderone}, G. and {Cano Rodr{\'\i}guez}, J.~L. and {Cara}, M. and {Cardoso}, J.~V.~M. and {Cheedella}, S. and {Copin}, Y. and {Corrales}, L. and {Crichton}, D. and {D'Avella}, D. and {Deil}, C. and {Depagne}, {\'E}. and {Dietrich}, J.~P. and {Donath}, A. and {Droettboom}, M. and {Earl}, N. and {Erben}, T. and {Fabbro}, S. and {Ferreira}, L.~A. and {Finethy}, T. and {Fox}, R.~T. and {Garrison}, L.~H. and {Gibbons}, S.~L.~J. and {Goldstein}, D.~A. and {Gommers}, R. and {Greco}, J.~P. and {Greenfield}, P. and {Groener}, A.~M. and {Grollier}, F. and {Hagen}, A. and {Hirst}, P. and {Homeier}, D. and {Horton}, A.~J. and {Hosseinzadeh}, G. and {Hu}, L. and {Hunkeler}, J.~S. and {Ivezi{\'c}}, {\v{Z}}. and {Jain}, A. and {Jenness}, T. and {Kanarek}, G. and {Kendrew}, S. and {Kern}, N.~S. and {Kerzendorf}, W.~E. and {Khvalko}, A. and {King}, J. and {Kirkby}, D. and {Kulkarni}, A.~M. and {Kumar}, A. and {Lee}, A. and {Lenz}, D. and {Littlefair}, S.~P. and {Ma}, Z. and {Macleod}, D.~M. and {Mastropietro}, M. and {McCully}, C. and {Montagnac}, S. and {Morris}, B.~M. and {Mueller}, M. and {Mumford}, S.~J. and {Muna}, D. and {Murphy}, N.~A. and {Nelson}, S. and {Nguyen}, G.~H. and {Ninan}, J.~P. and {N{\"o}the}, M. and {Ogaz}, S. and {Oh}, S. and {Parejko}, J.~K. and {Parley}, N. and {Pascual}, S. and {Patil}, R. and {Patil}, A.~A. and {Plunkett}, A.~L. and {Prochaska}, J.~X. and {Rastogi}, T. and {Reddy Janga}, V. and {Sabater}, J. and {Sakurikar}, P. and {Seifert}, M. and {Sherbert}, L.~E. and {Sherwood-Taylor}, H. and {Shih}, A.~Y. and {Sick}, J. and {Silbiger}, M.~T. and {Singanamalla}, S. and {Singer}, L.~P. and {Sladen}, P.~H. and {Sooley}, K.~A. and {Sornarajah}, S. and {Streicher}, O. and {Teuben}, P. and {Thomas}, S.~W. and {Tremblay}, G.~R. and {Turner}, J.~E.~H. and {Terr{\'o}n}, V. and {van Kerkwijk}, M.~H. and {de la Vega}, A. and {Watkins}, L.~L. and {Weaver}, B.~A. and {Whitmore}, J.~B. and {Woillez}, J. and {Zabalza}, V. and {Astropy Contributors}},
        title = "{The Astropy Project: Building an Open-science Project and Status of the v2.0 Core Package}",
      journal = {\aj},
         year = 2018,
        month = sep,
       volume = {156},
       number = {3},
          eid = {123},
        pages = {123},
          doi = {10.3847/1538-3881/aabc4f},
archivePrefix = {arXiv},
       eprint = {1801.02634},
 primaryClass = {astro-ph.IM},
       adsurl = {https://ui.adsabs.harvard.edu/abs/2018AJ....156..123A}
}

@ARTICLE{Astropy_13,
       author = {{Astropy Collaboration} and {Robitaille}, Thomas P. and {Tollerud}, Erik J. and {Greenfield}, Perry and {Droettboom}, Michael and {Bray}, Erik and {Aldcroft}, Tom and {Davis}, Matt and {Ginsburg}, Adam and {Price-Whelan}, Adrian M. and {Kerzendorf}, Wolfgang E. and {Conley}, Alexander and {Crighton}, Neil and {Barbary}, Kyle and {Muna}, Demitri and {Ferguson}, Henry and {Grollier}, Fr{\'e}d{\'e}ric and {Parikh}, Madhura M. and {Nair}, Prasanth H. and {Unther}, Hans M. and {Deil}, Christoph and {Woillez}, Julien and {Conseil}, Simon and {Kramer}, Roban and {Turner}, James E.~H. and {Singer}, Leo and {Fox}, Ryan and {Weaver}, Benjamin A. and {Zabalza}, Victor and {Edwards}, Zachary I. and {Azalee Bostroem}, K. and {Burke}, D.~J. and {Casey}, Andrew R. and {Crawford}, Steven M. and {Dencheva}, Nadia and {Ely}, Justin and {Jenness}, Tim and {Labrie}, Kathleen and {Lim}, Pey Lian and {Pierfederici}, Francesco and {Pontzen}, Andrew and {Ptak}, Andy and {Refsdal}, Brian and {Servillat}, Mathieu and {Streicher}, Ole},
        title = "{Astropy: A community Python package for astronomy}",
      journal = {\aap},
         year = 2013,
        month = oct,
       volume = {558},
          eid = {A33},
        pages = {A33},
          doi = {10.1051/0004-6361/201322068},
archivePrefix = {arXiv},
       eprint = {1307.6212},
 primaryClass = {astro-ph.IM},
       adsurl = {https://ui.adsabs.harvard.edu/abs/2013A&A...558A..33A}
}

@article{Harris20_numpy,
	title = {Array programming with {NumPy}},
	volume = {585},
	issn = {1476-4687},
	url = {https://doi.org/10.1038/s41586-020-2649-2},
	doi = {10.1038/s41586-020-2649-2},
	number = {7825},
	journal = {Nature},
	author = {Harris, Charles R. and Millman, K. Jarrod and van der Walt, Stéfan J. and Gommers, Ralf and Virtanen, Pauli and Cournapeau, David and Wieser, Eric and Taylor, Julian and Berg, Sebastian and Smith, Nathaniel J. and Kern, Robert and Picus, Matti and Hoyer, Stephan and van Kerkwijk, Marten H. and Brett, Matthew and Haldane, Allan and del Río, Jaime Fernández and Wiebe, Mark and Peterson, Pearu and Gérard-Marchant, Pierre and Sheppard, Kevin and Reddy, Tyler and Weckesser, Warren and Abbasi, Hameer and Gohlke, Christoph and Oliphant, Travis E.},
	month = sep,
	year = {2020},
	pages = {357--362},
}

@ARTICLE{Morii24,
       author = {{Morii}, Kaho and {Sanhueza}, Patricio and {Zhang}, Qizhou and {Nakamura}, Fumitaka and {Li}, Shanghuo and {Sabatini}, Giovanni and {Olguin}, Fernando A. and {Beuther}, Henrik and {Tafoya}, Daniel and {Izumi}, Natsuko and {Tatematsu}, Ken'ichi and {Sakai}, Takeshi},
        title = "{The ALMA Survey of 70 {\ensuremath{\mu}}m Dark High-mass Clumps in Early Stages (ASHES). XI. Statistical Study of Early Fragmentation}",
      journal = {\apj},
         year = 2024,
        month = may,
       volume = {966},
       number = {2},
          eid = {171},
        pages = {171},
          doi = {10.3847/1538-4357/ad32d0},
archivePrefix = {arXiv},
       eprint = {2403.07058},
 primaryClass = {astro-ph.GA},
       adsurl = {https://ui.adsabs.harvard.edu/abs/2024ApJ...966..171M}
}

@ARTICLE{Lu18,
       author = {{Lu}, Xing and {Zhang}, Qizhou and {Liu}, Hauyu Baobab and {Sanhueza}, Patricio and {Tatematsu}, Ken'ichi and {Feng}, Siyi and {Smith}, Howard A. and {Myers}, Philip C. and {Sridharan}, T.~K. and {Gu}, Qiusheng},
        title = "{Filamentary Fragmentation and Accretion in High-mass Star-forming Molecular Clouds}",
      journal = {\apj},
         year = 2018,
        month = mar,
       volume = {855},
       number = {1},
          eid = {9},
        pages = {9},
          doi = {10.3847/1538-4357/aaad11},
archivePrefix = {arXiv},
       eprint = {1801.05955},
 primaryClass = {astro-ph.GA},
       adsurl = {https://ui.adsabs.harvard.edu/abs/2018ApJ...855....9L}
}

@ARTICLE{Rygl13,
       author = {{Rygl}, K.~L.~J. and {Wyrowski}, F. and {Schuller}, F. and {Menten}, K.~M.},
        title = "{Initial phases of massive star formation in high infrared extinction clouds. II. Infall and onset of star formation}",
      journal = {\aap},
         year = 2013,
        month = jan,
       volume = {549},
          eid = {A5},
        pages = {A5},
          doi = {10.1051/0004-6361/201219574},
archivePrefix = {arXiv},
       eprint = {1210.2063},
 primaryClass = {astro-ph.GA},
       adsurl = {https://ui.adsabs.harvard.edu/abs/2013A&A...549A...5R}
}

@ARTICLE{Ishihara25,
       author = {{Ishihara}, Kousuke and {Nakamura}, Fumitaka and {Sanhueza}, Patricio and {Saito}, Masao},
        title = "{Turbulent fragmentation as the primary driver of core formation in Polaris Flare and Lupus I}",
      journal = {\aap},
         year = 2025,
        month = mar,
       volume = {695},
          eid = {L25},
        pages = {L25},
          doi = {10.1051/0004-6361/202452427},
archivePrefix = {arXiv},
       eprint = {2503.06613},
 primaryClass = {astro-ph.GA},
       adsurl = {https://ui.adsabs.harvard.edu/abs/2025A&A...695L..25I}
}

@ARTICLE{Gupta26,
       author = {{Gupta}, Shivani and {Baug}, Tapas and {Soam}, Archana and {Liu}, Tie and {Xu}, Fengwei and {Moharana}, Satyajeet and {Garay}, Guido and {Lee}, Chang Won and {Zhang}, Siju and {Hoque}, Ariful and {Porel}, Puja and {Zhu}, Lei and {Yang}, Dongting and {Liu}, HongLi and {Jiao}, Wenyu and {Liu}, Xunchuan and {Panja}, Alik and {Mai}, Xiaofeng and {Zhang}, Yankun and {Kim}, Shinyoung},
        title = "{ATOMS-QUARKS Survey: Inflow and Infall in Massive Protocluster G318.049+00.086{\textemdash} Evidence of Competitive Accretion}",
      journal = {\apj},
         year = 2026,
        month = mar,
       volume = {999},
       number = {2},
          eid = {180},
        pages = {180},
          doi = {10.3847/1538-4357/ae40fd},
archivePrefix = {arXiv},
       eprint = {2602.01238},
 primaryClass = {astro-ph.GA},
       adsurl = {https://ui.adsabs.harvard.edu/abs/2026ApJ...999..180G}
}

@ARTICLE{Xu23_hcn,
       author = {{Xu}, Fengwei and {Wang}, Ke and {He}, Yuxin and {Wu}, Jingwen and {Zhu}, Lei and {Mardones}, Diego},
        title = "{Clump-scale Gas Infall in High-mass Star Formation: A Multitransition View with James Clerk Maxwell Telescope HCN (4-3) Mapping}",
      journal = {\apjs},
         year = 2023,
        month = dec,
       volume = {269},
       number = {2},
          eid = {38},
        pages = {38},
          doi = {10.3847/1538-4365/acfee2},
archivePrefix = {arXiv},
       eprint = {2309.14686},
 primaryClass = {astro-ph.GA},
       adsurl = {https://ui.adsabs.harvard.edu/abs/2023ApJS..269...38X}
}

@ARTICLE{De_Vries05,
       author = {{De Vries}, Christopher H. and {Myers}, Philip C.},
        title = "{Molecular Line Profile Fitting with Analytic Radiative Transfer Models}",
      journal = {\apj},
         year = 2005,
        month = feb,
       volume = {620},
       number = {2},
        pages = {800-815},
          doi = {10.1086/427141},
archivePrefix = {arXiv},
       eprint = {astro-ph/0410748},
 primaryClass = {astro-ph},
       adsurl = {https://ui.adsabs.harvard.edu/abs/2005ApJ...620..800D}
}

@ARTICLE{Evans99,
       author = {{Evans}, Neal J.},
        title = "{Physical Conditions in Regions of Star Formation}",
      journal = {\araa},
         year = 1999,
        month = jan,
       volume = {37},
        pages = {311-362},
          doi = {10.1146/annurev.astro.37.1.311},
archivePrefix = {arXiv},
       eprint = {astro-ph/9905050},
 primaryClass = {astro-ph},
       adsurl = {https://ui.adsabs.harvard.edu/abs/1999ARA&A..37..311E}
}

@ARTICLE{Jin-Jin21,
       author = {{Xie}, Jin-Jin and {Wu}, Jing-Wen and {Fuller}, Gary A. and {Peretto}, Nicolas and {Ren}, Zhi-Yuan and {Chen}, Long-Fei and {Yan}, Yao-Ting and {Li}, Guo-Dong and {Duan}, Yan and {Xia}, Ji-Feng and {Wang}, Yong-Xiong and {Li}, Di},
        title = "{Studying infall in infrared dark clouds with multiple HCO$^{+}$ transitions}",
      journal = {Research in Astronomy and Astrophysics},
         year = 2021,
        month = oct,
       volume = {21},
       number = {8},
          eid = {208},
        pages = {208},
          doi = {10.1088/1674-4527/21/8/208},
archivePrefix = {arXiv},
       eprint = {2104.07231},
 primaryClass = {astro-ph.GA},
       adsurl = {https://ui.adsabs.harvard.edu/abs/2021RAA....21..208X}
}

@article{Smith09,
	adsurl = {https://ui.adsabs.harvard.edu/abs/2009MNRAS.400.1775S},
	archiveprefix = {arXiv},
	author = {{Smith}, Rowan J. and {Longmore}, Steven and {Bonnell}, Ian},
	doi = {10.1111/j.1365-2966.2009.15621.x},
	eprint = {0908.3910},
	journal = {\mnras},
	month = dec,
	number = {4},
	pages = {1775-1784},
	primaryclass = {astro-ph.SR},
	title = {{The simultaneous formation of massive stars and stellar clusters}},
	volume = {400},
	year = 2009}

@article{Redaelli22,
	adsurl = {https://ui.adsabs.harvard.edu/abs/2022ApJ...936..169R},
	author = {{Redaelli}, Elena and {Bovino}, Stefano and {Sanhueza}, Patricio and {Morii}, Kaho and {Sabatini}, Giovanni and {Caselli}, Paola and {Giannetti}, Andrea and {Li}, Shanghuo},
	doi = {10.3847/1538-4357/ac85b4},
	eid = {169},
	journal = {\apj},
	month = sep,
	number = {2},
	pages = {169},
	title = {{The Core Population and Kinematics of a Massive Clump at Early Stages: An Atacama Large Millimeter/submillimeter Array View}},
	volume = {936},
	year = 2022}

@ARTICLE{Li22,
       author = {{Li}, Shanghuo and {Sanhueza}, Patricio and {Lu}, Xing and {Lee}, Chang Won and {Zhang}, Qizhou and {Bovino}, Stefano and {Sabatini}, Giovanni and {Liu}, Tie and {Kim}, Kee-Tae and {Morii}, Kaho and {Tafoya}, Daniel and {Tatematsu}, Ken'ichi and {Sakai}, Takeshi and {Wang}, Junzhi and {Li}, Fei and {Silva}, Andrea and {Izumi}, Natsuko and {Allingham}, David},
        title = "{The ALMA Survey of 70 {\ensuremath{\mu}}m Dark High-mass Clumps in Early Stages (ASHES). VII. Chemistry of Embedded Dense Cores}",
      journal = {\apj},
         year = 2022,
        month = nov,
       volume = {939},
       number = {2},
          eid = {102},
        pages = {102},
          doi = {10.3847/1538-4357/ac94d4},
archivePrefix = {arXiv},
       eprint = {2209.12814},
 primaryClass = {astro-ph.GA},
       adsurl = {https://ui.adsabs.harvard.edu/abs/2022ApJ...939..102L}
}

@article{Sabatini22,
	adsurl = {https://ui.adsabs.harvard.edu/abs/2022ApJ...936...80S},
	archiveprefix = {arXiv},
	author = {{Sabatini}, Giovanni and {Bovino}, Stefano and {Sanhueza}, Patricio and {Morii}, Kaho and {Li}, Shanghuo and {Redaelli}, Elena and {Zhang}, Qizhou and {Lu}, Xing and {Feng}, Siyi and {Tafoya}, Daniel and {Izumi}, Natsuko and {Sakai}, Takeshi and {Tatematsu}, Ken'ichi and {Allingham}, David},
	doi = {10.3847/1538-4357/ac83aa},
	eid = {80},
	eprint = {2207.12431},
	journal = {\apj},
	month = sep,
	number = {1},
	pages = {80},
	primaryclass = {astro-ph.GA},
	title = {{The ALMA Survey of 70 {\ensuremath{\mu}}m Dark High-mass Clumps in Early Stages (ASHES). VI. The Core-scale CO Depletion}},
	volume = {936},
	year = 2022}

@article{Chambers09,
	adsurl = {https://ui.adsabs.harvard.edu/abs/2009ApJS..181..360C},
	author = {{Chambers}, E.~T. and {Jackson}, J.~M. and {Rathborne}, J.~M. and {Simon}, R.},
	doi = {10.1088/0067-0049/181/2/360},
	journal = {\apjs},
	month = apr,
	number = {2},
	pages = {360-390},
	title = {{Star Formation Activity of Cores within Infrared Dark Clouds}},
	volume = {181},
	year = 2009}

@ARTICLE{Gomez21,
       author = {{G{\'o}mez}, Gilberto C. and {V{\'a}zquez-Semadeni}, Enrique and {Palau}, Aina},
        title = "{Density profile evolution during prestellar core collapse: collapse starts at the large scale}",
      journal = {\mnras},
         year = 2021,
        month = apr,
       volume = {502},
       number = {4},
        pages = {4963-4971},
          doi = {10.1093/mnras/stab394},
archivePrefix = {arXiv},
       eprint = {2009.14151},
 primaryClass = {astro-ph.GA},
       adsurl = {https://ui.adsabs.harvard.edu/abs/2021MNRAS.502.4963G}
}

@inproceedings{Evans03,
	adsurl = {https://ui.adsabs.harvard.edu/abs/2003cdsf.conf..157E},
	archiveprefix = {arXiv},
	author = {{Evans}, Neal. J.},
	booktitle = {SFChem 2002: Chemistry as a Diagnostic of Star Formation},
	editor = {{Curry}, Charles L. and {Fich}, Michel},
	eprint = {astro-ph/0211526},
	month = jan,
	pages = {157},
	primaryclass = {astro-ph},
	title = {{Studying Infall}},
	year = 2003}

@article{Sakai21,
	adsurl = {https://ui.adsabs.harvard.edu/abs/2022ApJ...925..144S},
	archiveprefix = {arXiv},
	author = {{Sakai}, Takeshi and {Sanhueza}, Patricio and {Furuya}, Kenji and {Tatematsu}, Ken'ichi and {Li}, Shanghuo and {Aikawa}, Yuri and {Lu}, Xing and {Zhang}, Qizhou and {Morii}, Kaho and {Nakamura}, Fumitaka and {Takemura}, Hideaki and {Izumi}, Natsuko and {Hirota}, Tomoya and {Silva}, Andrea and {Guzman}, Andres E. and {Sakai}, Nami and {Yamamoto}, Satoshi},
	doi = {10.3847/1538-4357/ac3d2e},
	eid = {144},
	eprint = {2111.13325},
	journal = {\apj},
	month = feb,
	number = {2},
	pages = {144},
	primaryclass = {astro-ph.GA},
	title = {{The ALMA Survey of 70 {\ensuremath{\mu}}m Dark High-mass Clumps in Early Stages (ASHES). V. Deuterated Molecules in the 70 {\ensuremath{\mu}}m Dark IRDC G14.492-00.139}},
	volume = {925},
	year = 2022}

@ARTICLE{Sanhueza25,
       author = {{Sanhueza}, Patricio and {Liu}, Junhao and {Morii}, Kaho and {Girart}, Josep Miquel and {Zhang}, Qizhou and {Stephens}, Ian W. and {Jackson}, James M. and {Cort{\'e}s}, Paulo C. and {Koch}, Patrick M. and {Cyganowski}, Claudia J. and {Saha}, Piyali and {Beuther}, Henrik and {Zhang}, Suinan and {Beltr{\'a}n}, Maria T. and {Cheng}, Yu and {Olguin}, Fernando A. and {Lu}, Xing and {Choudhury}, Spandan and {Pattle}, Kate and {Fern{\'a}ndez-L{\'o}pez}, Manuel and {Hwang}, Jihye and {Kang}, Ji-hyun and {Karoly}, Janik and {Ginsburg}, Adam and {Lyo}, A. -Ran and {Taniguchi}, Kotomi and {Jiao}, Wenyu and {Eswaraiah}, Chakali and {Luo}, Qiu-yi and {Wang}, Jia-Wei and {Commer{\c{c}}on}, Beno{\^\i}t and {Li}, Shanghuo and {Xu}, Fengwei and {Chen}, Huei-Ru Vivien and {Zapata}, Luis A. and {Chung}, Eun Jung and {Nakamura}, Fumitaka and {Panigrahy}, Sandhyarani and {Sakai}, Takeshi},
        title = "{Magnetic Fields in Massive Star-forming Regions (MagMaR). V. The Magnetic Field at the Onset of High-mass Star Formation}",
      journal = {\apj},
         year = 2025,
        month = feb,
       volume = {980},
       number = {1},
          eid = {87},
        pages = {87},
          doi = {10.3847/1538-4357/ad9d40},
archivePrefix = {arXiv},
       eprint = {2412.08790},
 primaryClass = {astro-ph.GA},
       adsurl = {https://ui.adsabs.harvard.edu/abs/2025ApJ...980...87S}
}

@article{Traficante17,
	adsurl = {https://ui.adsabs.harvard.edu/abs/2017MNRAS.470.3882T},
	archiveprefix = {arXiv},
	author = {{Traficante}, A. and {Fuller}, G.~A. and {Billot}, N. and {Duarte-Cabral}, A. and {Merello}, M. and {Molinari}, S. and {Peretto}, N. and {Schisano}, E.},
	doi = {10.1093/mnras/stx1375},
	eprint = {1706.00432},
	journal = {\mnras},
	month = oct,
	number = {4},
	pages = {3882-3923},
	primaryclass = {astro-ph.GA},
	title = {{Massive 70 {\ensuremath{\mu}}m quiet clumps I: evidence of embedded low/intermediate-mass star formation activity}},
	volume = {470},
	year = 2017}

@article{Reiter11,
	adsurl = {https://ui.adsabs.harvard.edu/abs/2011ApJ...740...40R},
	archiveprefix = {arXiv},
	author = {{Reiter}, Megan and {Shirley}, Yancy L. and {Wu}, Jingwen and {Brogan}, Crystal and {Wootten}, Alwyn and {Tatematsu}, Ken'ichi},
	doi = {10.1088/0004-637X/740/1/40},
	eid = {40},
	eprint = {1107.5343},
	journal = {\apj},
	month = oct,
	number = {1},
	pages = {40},
	primaryclass = {astro-ph.SR},
	title = {{Evidence for Inflow in High-mass Star-forming Clumps}},
	volume = {740},
	year = 2011}

@article{Smith13,
	adsurl = {https://ui.adsabs.harvard.edu/abs/2013ApJ...771...24S},
	archiveprefix = {arXiv},
	author = {{Smith}, Rowan J. and {Shetty}, Rahul and {Beuther}, Henrik and {Klessen}, Ralf S. and {Bonnell}, Ian A.},
	doi = {10.1088/0004-637X/771/1/24},
	eid = {24},
	eprint = {1304.4950},
	journal = {\apj},
	month = jul,
	number = {1},
	pages = {24},
	primaryclass = {astro-ph.GA},
	title = {{Line Profiles of Cores within Clusters. II. Signatures of Dynamical Collapse during High-mass Star Formation}},
	volume = {771},
	year = 2013}

@article{Smith12,
	adsurl = {https://ui.adsabs.harvard.edu/abs/2012ApJ...750...64S},
	archiveprefix = {arXiv},
	author = {{Smith}, Rowan J. and {Shetty}, Rahul and {Stutz}, Amelia M. and {Klessen}, Ralf S.},
	doi = {10.1088/0004-637X/750/1/64},
	eid = {64},
	eprint = {1201.6275},
	journal = {\apj},
	month = may,
	number = {1},
	pages = {64},
	primaryclass = {astro-ph.GA},
	title = {{Line Profiles of Cores within Clusters. I. The Anatomy of a Filament}},
	volume = {750},
	year = 2012}

@article{Campbell16,
	adsurl = {https://ui.adsabs.harvard.edu/abs/2016ApJ...819..143C},
	archiveprefix = {arXiv},
	author = {{Campbell}, J.~L. and {Friesen}, R.~K. and {Martin}, P.~G. and {Caselli}, P. and {Kauffmann}, J. and {Pineda}, J.~E.},
	doi = {10.3847/0004-637X/819/2/143},
	eid = {143},
	eprint = {1601.07165},
	journal = {\apj},
	month = mar,
	number = {2},
	pages = {143},
	primaryclass = {astro-ph.SR},
	title = {{Contraction Signatures toward Dense Cores in the Perseus Molecular Cloud}},
	volume = {819},
	year = 2016}

@article{Jackson19,
	adsurl = {https://ui.adsabs.harvard.edu/abs/2019ApJ...870....5J},
	archiveprefix = {arXiv},
	author = {{Jackson}, James M. and {Whitaker}, J. Scott and {Rathborne}, J.~M. and {Foster}, J.~B. and {Contreras}, Y. and {Sanhueza}, Patricio and {Stephens}, Ian W. and {Longmore}, S.~N. and {Allingham}, David},
	doi = {10.3847/1538-4357/aaef84},
	eid = {5},
	eprint = {1811.05545},
	journal = {\apj},
	month = jan,
	number = {1},
	pages = {5},
	primaryclass = {astro-ph.GA},
	title = {{Asymmetric Line Profiles in Dense Molecular Clumps Observed in MALT90: Evidence for Global Collapse}},
	volume = {870},
	year = 2019}

@article{Chira14,
	adsurl = {https://ui.adsabs.harvard.edu/abs/2014MNRAS.444..874C},
	archiveprefix = {arXiv},
	author = {{Chira}, Roxana-Adela and {Smith}, Rowan J. and {Klessen}, Ralf S. and {Stutz}, Amelia M. and {Shetty}, Rahul},
	doi = {10.1093/mnras/stu1497},
	eprint = {1402.5279},
	journal = {\mnras},
	month = oct,
	number = {1},
	pages = {874-886},
	primaryclass = {astro-ph.SR},
	title = {{Line profiles of cores within clusters - III. What is the most reliable tracer of core collapse in dense clusters?}},
	volume = {444},
	year = 2014}

@article{Wyrowski16,
	adsurl = {https://ui.adsabs.harvard.edu/abs/2016A&A...585A.149W},
	archiveprefix = {arXiv},
	author = {{Wyrowski}, F. and {G{\"u}sten}, R. and {Menten}, K.~M. and {Wiesemeyer}, H. and {Csengeri}, T. and {Heyminck}, S. and {Klein}, B. and {K{\"o}nig}, C. and {Urquhart}, J.~S.},
	doi = {10.1051/0004-6361/201526361},
	eid = {A149},
	eprint = {1510.08374},
	journal = {\aap},
	month = jan,
	pages = {A149},
	primaryclass = {astro-ph.SR},
	title = {{Infall through the evolution of high-mass star-forming clumps}},
	volume = {585},
	year = 2016}

@article{Yang21,
	adsurl = {https://ui.adsabs.harvard.edu/abs/2021ApJ...922..144Y},
	archiveprefix = {arXiv},
	author = {{Yang}, Yang and {Jiang}, Zhibo and {Chen}, Zhiwei and {Ao}, Yiping and {Yu}, Shuling},
	doi = {10.3847/1538-4357/ac22ab},
	eid = {144},
	eprint = {2109.07221},
	journal = {\apj},
	month = dec,
	number = {2},
	pages = {144},
	primaryclass = {astro-ph.GA},
	title = {{In Search of Infall Motion in Molecular Clumps. III. HCO+ (1-0) and H13CO+ (1-0) Mapping Observations toward Confirmed Infall Sources}},
	volume = {922},
	year = 2021}

@article{Morii21,
	adsurl = {https://ui.adsabs.harvard.edu/abs/2021ApJ...923..147M},
	archiveprefix = {arXiv},
	author = {{Morii}, Kaho and {Sanhueza}, Patricio and {Nakamura}, Fumitaka and {Jackson}, James M. and {Li}, Shanghuo and {Beuther}, Henrik and {Zhang}, Qizhou and {Feng}, Siyi and {Tafoya}, Daniel and {Guzm{\'a}n}, Andr{\'e}s E. and {Izumi}, Natsuko and {Sakai}, Takeshi and {Lu}, Xing and {Tatematsu}, Ken'ichi and {Ohashi}, Satoshi and {Silva}, Andrea and {Olguin}, Fernando A. and {Contreras}, Yanett},
	doi = {10.3847/1538-4357/ac2365},
	eid = {147},
	eprint = {2109.01231},
	journal = {\apj},
	month = dec,
	number = {2},
	pages = {147},
	primaryclass = {astro-ph.GA},
	title = {{The ALMA Survey of 70 {\ensuremath{\mu}}m Dark High-mass Clumps in Early Stages (ASHES). IV. Star Formation Signatures in G023.477}},
	volume = {923},
	year = 2021}

@article{Wang10,
	adsurl = {https://ui.adsabs.harvard.edu/abs/2010ApJ...709...27W},
	archiveprefix = {arXiv},
	author = {{Wang}, Peng and {Li}, Zhi-Yun and {Abel}, Tom and {Nakamura}, Fumitaka},
	doi = {10.1088/0004-637X/709/1/27},
	eprint = {0908.4129},
	journal = {\apj},
	month = jan,
	number = {1},
	pages = {27-41},
	primaryclass = {astro-ph.SR},
	title = {{Outflow Feedback Regulated Massive Star Formation in Parsec-Scale Cluster-Forming Clumps}},
	volume = {709},
	year = 2010}

@article{Foster11,
	adsurl = {https://ui.adsabs.harvard.edu/abs/2011ApJS..197...25F},
	archiveprefix = {arXiv},
	author = {{Foster}, Jonathan B. and {Jackson}, James M. and {Barnes}, Peter J. and {Barris}, Elizabeth and {Brooks}, Kate and {Cunningham}, Maria and {Finn}, Susanna C. and {Fuller}, Gary A. and {Longmore}, Steve N. and {Mascoop}, Joshua L. and {Peretto}, Nicolas and {Rathborne}, Jill and {Sanhueza}, Patricio and {Schuller}, Fr{\'e}d{\'e}ric and {Wyrowski}, Friedrich},
	doi = {10.1088/0067-0049/197/2/25},
	eid = {25},
	eprint = {1108.4446},
	journal = {\apjs},
	month = dec,
	number = {2},
	pages = {25},
	primaryclass = {astro-ph.GA},
	title = {{The Millimeter Astronomy Legacy Team 90 GHz (MALT90) Pilot Survey}},
	volume = {197},
	year = 2011}

@article{Whitaker17,
	adsurl = {https://ui.adsabs.harvard.edu/abs/2017AJ....154..140W},
	author = {{Whitaker}, J. Scott and {Jackson}, James M. and {Rathborne}, J.~M. and {Foster}, J.~B. and {Contreras}, Y. and {Sanhueza}, Patricio and {Stephens}, Ian W. and {Longmore}, S.~N.},
	doi = {10.3847/1538-3881/aa86ad},
	eid = {140},
	journal = {\aj},
	month = oct,
	number = {4},
	pages = {140},
	title = {{MALT90 Kinematic Distances to Dense Molecular Clumps}},
	volume = {154},
	year = 2017}

@article{Sanhueza21,
	adsurl = {https://ui.adsabs.harvard.edu/abs/2021ApJ...915L..10S},
	archiveprefix = {arXiv},
	author = {{Sanhueza}, Patricio and {Girart}, Josep Miquel and {Padovani}, Marco and {Galli}, Daniele and {Hull}, Charles L.~H. and {Zhang}, Qizhou and {Cortes}, Paulo and {Stephens}, Ian W. and {Fern{\'a}ndez-L{\'o}pez}, Manuel and {Jackson}, James M. and {Frau}, Pau and {Kock}, Patrick M. and {Wu}, Benjamin and {Zapata}, Luis A. and {Olguin}, Fernando and {Lu}, Xing and {Silva}, Andrea and {Tang}, Ya-Wen and {Sakai}, Takeshi and {Guzm{\'a}n}, Andr{\'e}s E. and {Tatematsu}, Ken'ichi and {Nakamura}, Fumitaka and {Chen}, Huei-Ru Vivien},
	doi = {10.3847/2041-8213/ac081c},
	eid = {L10},
	eprint = {2106.03866},
	journal = {\apjl},
	month = jul,
	number = {1},
	pages = {L10},
	primaryclass = {astro-ph.GA},
	title = {{Gravity-driven Magnetic Field at 1000 au Scales in High-mass Star Formation}},
	volume = {915},
	year = 2021}

@inproceedings{Menten05,
	adsurl = {https://ui.adsabs.harvard.edu/abs/2005IAUS..227...23M},
	archiveprefix = {arXiv},
	author = {{Menten}, K.~M. and {Pillai}, T. and {Wyrowski}, F.},
	booktitle = {Massive Star Birth: A Crossroads of Astrophysics},
	doi = {10.1017/S1743921305004321},
	editor = {{Cesaroni}, R. and {Felli}, M. and {Churchwell}, E. and {Walmsley}, M.},
	eprint = {astro-ph/0508030},
	month = jan,
	pages = {23-34},
	primaryclass = {astro-ph},
	title = {{Initial conditions for massive star birth-Infrared dark clouds}},
	volume = {227},
	year = 2005}

@article{Jackson13,
	adsurl = {https://ui.adsabs.harvard.edu/abs/2013PASA...30...57J},
	archiveprefix = {arXiv},
	author = {{Jackson}, J.~M. and {Rathborne}, J.~M. and {Foster}, J.~B. and {Whitaker}, J.~S. and {Sanhueza}, P. and {Claysmith}, C. and {Mascoop}, J.~L. and {Wienen}, M. and {Breen}, S.~L. and {Herpin}, F. and {Duarte-Cabral}, A. and {Csengeri}, T. and {Longmore}, S.~N. and {Contreras}, Y. and {Indermuehle}, B. and {Barnes}, P.~J. and {Walsh}, A.~J. and {Cunningham}, M.~R. and {Brooks}, K.~J. and {Britton}, T.~R. and {Voronkov}, M.~A. and {Urquhart}, J.~S. and {Alves}, J. and {Jordan}, C.~H. and {Hill}, T. and {Hoq}, S. and {Finn}, S.~C. and {Bains}, I. and {Bontemps}, S. and {Bronfman}, L. and {Caswell}, J.~L. and {Deharveng}, L. and {Ellingsen}, S.~P. and {Fuller}, G.~A. and {Garay}, G. and {Green}, J.~A. and {Hindson}, L. and {Jones}, P.~A. and {Lenfestey}, C. and {Lo}, N. and {Lowe}, V. and {Mardones}, D. and {Menten}, K.~M. and {Minier}, V. and {Morgan}, L.~K. and {Motte}, F. and {Muller}, E. and {Peretto}, N. and {Purcell}, C.~R. and {Schilke}, P. and {Bontemps}, Schneider-N. and {Schuller}, F. and {Titmarsh}, A. and {Wyrowski}, F. and {Zavagno}, A.},
	doi = {10.1017/pasa.2013.37},
	eid = {e057},
	eprint = {1310.1131},
	journal = {\pasa},
	month = nov,
	pages = {e057},
	primaryclass = {astro-ph.GA},
	title = {{MALT90: The Millimetre Astronomy Legacy Team 90 GHz Survey}},
	volume = {30},
	year = 2013}

@article{Tafoya21,
	adsurl = {https://ui.adsabs.harvard.edu/abs/2021ApJ...913..131T},
	archiveprefix = {arXiv},
	author = {{Tafoya}, Daniel and {Sanhueza}, Patricio and {Zhang}, Qizhou and {Li}, Shanghuo and {Guzm{\'a}n}, Andr{\'e}s E. and {Silva}, Andrea and {de la Fuente}, Eduardo and {Lu}, Xing and {Morii}, Kaho and {Tatematsu}, Ken'ichi and {Contreras}, Yanett and {Izumi}, Natsuko and {Jackson}, James M. and {Nakamura}, Fumitaka and {Sakai}, Takeshi},
	doi = {10.3847/1538-4357/abf5da},
	eid = {131},
	eprint = {2104.02625},
	journal = {\apj},
	month = jun,
	number = {2},
	pages = {131},
	primaryclass = {astro-ph.GA},
	title = {{The ALMA Survey of 70 {\ensuremath{\mu}}m Dark High-mass Clumps in Early Stages (ASHES). III. A Young Molecular Outflow Driven by a Decelerating Jet}},
	volume = {913},
	year = 2021}

@article{Csengeri11,
	adsurl = {https://ui.adsabs.harvard.edu/abs/2011A&A...527A.135C},
	archiveprefix = {arXiv},
	author = {{Csengeri}, T. and {Bontemps}, S. and {Schneider}, N. and {Motte}, F. and {Dib}, S.},
	doi = {10.1051/0004-6361/201014984},
	eid = {A135},
	eprint = {1009.0598},
	journal = {\aap},
	month = mar,
	pages = {A135},
	primaryclass = {astro-ph.GA},
	title = {{Gas dynamics in massive dense cores in Cygnus-X}},
	volume = {527},
	year = 2011}

@article{Schneider10,
	adsurl = {https://ui.adsabs.harvard.edu/abs/2010A&A...520A..49S},
	archiveprefix = {arXiv},
	author = {{Schneider}, N. and {Csengeri}, T. and {Bontemps}, S. and {Motte}, F. and {Simon}, R. and {Hennebelle}, P. and {Federrath}, C. and {Klessen}, R.},
	doi = {10.1051/0004-6361/201014481},
	eid = {A49},
	eprint = {1003.4198},
	journal = {\aap},
	month = sep,
	pages = {A49},
	primaryclass = {astro-ph.GA},
	title = {{Dynamic star formation in the massive DR21 filament}},
	volume = {520},
	year = 2010}

@article{Henshaw14,
	adsurl = {https://ui.adsabs.harvard.edu/abs/2014MNRAS.440.2860H},
	archiveprefix = {arXiv},
	author = {{Henshaw}, J.~D. and {Caselli}, P. and {Fontani}, F. and {Jim{\'e}nez-Serra}, I. and {Tan}, J.~C.},
	doi = {10.1093/mnras/stu446},
	eprint = {1403.1444},
	journal = {\mnras},
	month = may,
	number = {3},
	pages = {2860-2881},
	primaryclass = {astro-ph.SR},
	title = {{The dynamical properties of dense filaments in the infrared dark cloud G035.39-00.33}},
	volume = {440},
	year = 2014}

@article{Sanhueza13,
	adsurl = {https://ui.adsabs.harvard.edu/abs/2013ApJ...773..123S},
	archiveprefix = {arXiv},
	author = {{Sanhueza}, Patricio and {Jackson}, James M. and {Foster}, Jonathan B. and {Jimenez-Serra}, Izaskun and {Dirienzo}, William J. and {Pillai}, Thushara},
	doi = {10.1088/0004-637X/773/2/123},
	eid = {123},
	eprint = {1307.1474},
	journal = {\apj},
	month = aug,
	number = {2},
	pages = {123},
	primaryclass = {astro-ph.GA},
	title = {{Distinct Chemical Regions in the ``Prestellar'' Infrared Dark Cloud G028.23-00.19}},
	volume = {773},
	year = 2013}

@article{Sanhueza12,
	adsurl = {https://ui.adsabs.harvard.edu/abs/2012ApJ...756...60S},
	archiveprefix = {arXiv},
	author = {{Sanhueza}, Patricio and {Jackson}, James M. and {Foster}, Jonathan B. and {Garay}, Guido and {Silva}, Andrea and {Finn}, Susanna C.},
	doi = {10.1088/0004-637X/756/1/60},
	eid = {60},
	eprint = {1206.6500},
	journal = {\apj},
	month = sep,
	number = {1},
	pages = {60},
	primaryclass = {astro-ph.GA},
	title = {{Chemistry in Infrared Dark Cloud Clumps: A Molecular Line Survey at 3 mm}},
	volume = {756},
	year = 2012}

@article{Sanhueza10,
	adsurl = {https://ui.adsabs.harvard.edu/abs/2010ApJ...715...18S},
	archiveprefix = {arXiv},
	author = {{Sanhueza}, P. and {Garay}, G. and {Bronfman}, L. and {Mardones}, D. and {May}, J. and {Saito}, M.},
	doi = {10.1088/0004-637X/715/1/18},
	eprint = {1005.3048},
	journal = {\apj},
	month = may,
	number = {1},
	pages = {18-32},
	primaryclass = {astro-ph.GA},
	title = {{Molecular Outflows Within the Filamentary Infrared Dark Cloud G34.43+0.24}},
	volume = {715},
	year = 2010}

@article{Wilson94,
	adsurl = {https://ui.adsabs.harvard.edu/abs/1994ARA&A..32..191W},
	author = {{Wilson}, T.~L. and {Rood}, R.},
	doi = {10.1146/annurev.aa.32.090194.001203},
	journal = {\araa},
	month = jan,
	pages = {191-226},
	title = {{Abundances in the Interstellar Medium}},
	volume = {32},
	year = 1994}

@article{Chen19,
	adsurl = {https://ui.adsabs.harvard.edu/abs/2019ApJ...875...24C},
	archiveprefix = {arXiv},
	author = {{Chen}, Huei-Ru Vivien and {Zhang}, Qizhou and {Wright}, M.~C.~H. and {Busquet}, Gemma and {Lin}, Yuxin and {Liu}, Hauyu Baobab and {Olguin}, F.~A. and {Sanhueza}, Patricio and {Nakamura}, Fumitaka and {Palau}, Aina and {Ohashi}, Satoshi and {Tatematsu}, Ken'ichi and {Liao}, Li-Wen},
	doi = {10.3847/1538-4357/ab0f3e},
	eid = {24},
	eprint = {1903.04376},
	journal = {\apj},
	month = apr,
	number = {1},
	pages = {24},
	primaryclass = {astro-ph.GA},
	title = {{Filamentary Accretion Flows in the Infrared Dark Cloud G14.225-0.506 Revealed by ALMA}},
	volume = {875},
	year = 2019}

@article{Li20,
	adsurl = {https://ui.adsabs.harvard.edu/abs/2020ApJ...903..119L},
	archiveprefix = {arXiv},
	author = {{Li}, Shanghuo and {Sanhueza}, Patricio and {Zhang}, Qizhou and {Nakamura}, Fumitaka and {Lu}, Xing and {Wang}, Junzhi and {Liu}, Tie and {Tatematsu}, Ken'ichi and {Jackson}, James M. and {Silva}, Andrea and {Guzm{\'a}n}, Andr{\'e}s E. and {Sakai}, Takeshi and {Izumi}, Natsuko and {Tafoya}, Daniel and {Li}, Fei and {Contreras}, Yanett and {Morii}, Kaho and {Kim}, Kee-Tae},
	doi = {10.3847/1538-4357/abb81f},
	eid = {119},
	eprint = {2009.05506},
	journal = {\apj},
	month = nov,
	number = {2},
	pages = {119},
	primaryclass = {astro-ph.GA},
	title = {{The ALMA Survey of 70 {\ensuremath{\mu}}m Dark High-mass Clumps in Early Stages (ASHES). II. Molecular Outflows in the Extreme Early Stages of Protocluster Formation}},
	volume = {903},
	year = 2020}

@article{Padoan20,
	adsurl = {https://ui.adsabs.harvard.edu/abs/2020ApJ...900...82P},
	archiveprefix = {arXiv},
	author = {{Padoan}, Paolo and {Pan}, Liubin and {Juvela}, Mika and {Haugb{\o}lle}, Troels and {Nordlund}, {\r{A}}ke},
	doi = {10.3847/1538-4357/abaa47},
	eid = {82},
	eprint = {1911.04465},
	journal = {\apj},
	month = sep,
	number = {1},
	pages = {82},
	primaryclass = {astro-ph.GA},
	title = {{The Origin of Massive Stars: The Inertial-inflow Model}},
	volume = {900},
	year = 2020}

@article{vazquez19,
	adsurl = {https://ui.adsabs.harvard.edu/abs/2019MNRAS.490.3061V},
	archiveprefix = {arXiv},
	author = {{V{\'a}zquez-Semadeni}, Enrique and {Palau}, Aina and {Ballesteros-Paredes}, Javier and {G{\'o}mez}, Gilberto C. and {Zamora-Avil{\'e}s}, Manuel},
	doi = {10.1093/mnras/stz2736},
	eprint = {1903.11247},
	journal = {\mnras},
	month = dec,
	number = {3},
	pages = {3061-3097},
	primaryclass = {astro-ph.GA},
	title = {{Global hierarchical collapse in molecular clouds. Towards a comprehensive scenario}},
	volume = {490},
	year = 2019}

@article{Rathborne06,
	adsurl = {https://ui.adsabs.harvard.edu/abs/2006ApJ...641..389R},
	archiveprefix = {arXiv},
	author = {{Rathborne}, J.~M. and {Jackson}, J.~M. and {Simon}, R.},
	doi = {10.1086/500423},
	eprint = {astro-ph/0602246},
	journal = {\apj},
	month = apr,
	number = {1},
	pages = {389-405},
	primaryclass = {astro-ph},
	title = {{Infrared Dark Clouds: Precursors to Star Clusters}},
	volume = {641},
	year = 2006}

@article{Contreras18,
	adsurl = {https://ui.adsabs.harvard.edu/abs/2018ApJ...861...14C},
	archiveprefix = {arXiv},
	author = {{Contreras}, Yanett and {Sanhueza}, Patricio and {Jackson}, James M. and {Guzm{\'a}n}, Andr{\'e}s E. and {Longmore}, Steven and {Garay}, Guido and {Zhang}, Qizhou and {Nguyễn-Lu'o'ng}, Quang and {Tatematsu}, Ken'ichi and {Nakamura}, Fumitaka and {Sakai}, Takeshi and {Ohashi}, Satoshi and {Liu}, Tie and {Saito}, Masao and {Gomez}, Laura and {Rathborne}, Jill and {Whitaker}, Scott},
	doi = {10.3847/1538-4357/aac2ec},
	eid = {14},
	eprint = {1805.01802},
	journal = {\apj},
	month = jul,
	number = {1},
	pages = {14},
	primaryclass = {astro-ph.GA},
	title = {{Infall Signatures in a Prestellar Core Embedded in the High-mass 70 {\ensuremath{\mu}}m Dark IRDC G331.372-00.116}},
	volume = {861},
	year = 2018}

@article{Bonnell04,
	adsurl = {https://ui.adsabs.harvard.edu/abs/2004MNRAS.349..735B},
	archiveprefix = {arXiv},
	author = {{Bonnell}, Ian A. and {Vine}, Stephen G. and {Bate}, Matthew R.},
	doi = {10.1111/j.1365-2966.2004.07543.x},
	eprint = {astro-ph/0401059},
	journal = {\mnras},
	month = apr,
	number = {2},
	pages = {735-741},
	primaryclass = {astro-ph},
	title = {{Massive star formation: nurture, not nature}},
	volume = {349},
	year = 2004}

@article{Sanhueza17,
	adsurl = {https://ui.adsabs.harvard.edu/abs/2017ApJ...841...97S},
	archiveprefix = {arXiv},
	author = {{Sanhueza}, Patricio and {Jackson}, James M. and {Zhang}, Qizhou and {Guzm{\'a}n}, Andr{\'e}s E. and {Lu}, Xing and {Stephens}, Ian W. and {Wang}, Ke and {Tatematsu}, Ken'ichi},
	doi = {10.3847/1538-4357/aa6ff8},
	eid = {97},
	eprint = {1704.08264},
	journal = {\apj},
	month = jun,
	number = {2},
	pages = {97},
	primaryclass = {astro-ph.GA},
	title = {{A Massive Prestellar Clump Hosting No High-mass Cores}},
	volume = {841},
	year = 2017}

@article{Sanhueza19,
	adsurl = {https://ui.adsabs.harvard.edu/abs/2019ApJ...886..102S},
	archiveprefix = {arXiv},
	author = {{Sanhueza}, Patricio and {Contreras}, Yanett and {Wu}, Benjamin and {Jackson}, James M. and {Guzm{\'a}n}, Andr{\'e}s E. and {Zhang}, Qizhou and {Li}, Shanghuo and {Lu}, Xing and {Silva}, Andrea and {Izumi}, Natsuko and {Liu}, Tie and {Miura}, Rie E. and {Tatematsu}, Ken{\textquoteright}ichi and {Sakai}, Takeshi and {Beuther}, Henrik and {Garay}, Guido and {Ohashi}, Satoshi and {Saito}, Masao and {Nakamura}, Fumitaka and {Saigo}, Kazuya and {Veena}, V.~S. and {Nguyen-Luong}, Quang and {Tafoya}, Daniel},
	doi = {10.3847/1538-4357/ab45e9},
	eid = {102},
	eprint = {1909.07985},
	journal = {\apj},
	month = dec,
	number = {2},
	pages = {102},
	primaryclass = {astro-ph.GA},
	title = {{The ALMA Survey of 70 {\ensuremath{\mu}}m Dark High-mass Clumps in Early Stages (ASHES). I. Pilot Survey: Clump Fragmentation}},
	volume = {886},
	year = 2019}

@article{Beuther13,
	adsurl = {https://ui.adsabs.harvard.edu/abs/2013A&A...553A.115B},
	archiveprefix = {arXiv},
	author = {{Beuther}, H. and {Linz}, H. and {Tackenberg}, J. and {Henning}, Th. and {Krause}, O. and {Ragan}, S. and {Nielbock}, M. and {Launhardt}, R. and {Bihr}, S. and {Schmiedeke}, A. and {Smith}, R. and {Sakai}, T.},
	doi = {10.1051/0004-6361/201220475},
	eid = {A115},
	eprint = {1304.6820},
	journal = {\aap},
	month = may,
	pages = {A115},
	primaryclass = {astro-ph.GA},
	title = {{Fragmentation and dynamical collapse of the starless high-mass star-forming region IRDC 18310-4}},
	volume = {553},
	year = 2013}

@article{Sridharan05,
	adsurl = {https://ui.adsabs.harvard.edu/abs/2005ApJ...634L..57S},
	archiveprefix = {arXiv},
	author = {{Sridharan}, T.~K. and {Beuther}, H. and {Saito}, M. and {Wyrowski}, F. and {Schilke}, P.},
	doi = {10.1086/498644},
	eprint = {astro-ph/0508421},
	journal = {\apjl},
	month = nov,
	number = {1},
	pages = {L57-L60},
	primaryclass = {astro-ph},
	title = {{High-Mass Starless Cores}},
	volume = {634},
	year = 2005}

@ARTICLE{Morii23,
       author = {{Morii}, Kaho and {Sanhueza}, Patricio and {Nakamura}, Fumitaka and {Zhang}, Qizhou and {Sabatini}, Giovanni and {Beuther}, Henrik and {Lu}, Xing and {Li}, Shanghuo and {Garay}, Guido and {Jackson}, James M. and {Olguin}, Fernando A. and {Tafoya}, Daniel and {Tatematsu}, Ken'ichi and {Izumi}, Natsuko and {Sakai}, Takeshi and {Silva}, Andrea},
        title = "{The ALMA Survey of 70 {\ensuremath{\mu}}m Dark High-mass Clumps in Early Stages (ASHES). IX. Physical Properties and Spatial Distribution of Cores in IRDCs}",
      journal = {\apj},
         year = 2023,
        month = jun,
       volume = {950},
       number = {2},
          eid = {148},
        pages = {148},
          doi = {10.3847/1538-4357/acccea},
archivePrefix = {arXiv},
       eprint = {2304.01757},
 primaryClass = {astro-ph.GA},
       adsurl = {https://ui.adsabs.harvard.edu/abs/2023ApJ...950..148M}
}

@ARTICLE{Li23,
       author = {{Li}, Shanghuo and {Sanhueza}, Patricio and {Zhang}, Qizhou and {Guido}, Garay and {Sabatini}, Giovanni and {Morii}, Kaho and {Lu}, Xing and {Tafoya}, Daniel and {Nakamura}, Fumitaka and {Izumi}, Natsuko and {Tatematsu}, Ken'ichi and {Li}, Fei},
        title = "{The ALMA Survey of 70 {\ensuremath{\mu}}m Dark High-mass Clumps in Early Stages (ASHES). VIII. Dynamics of Embedded Dense Cores}",
      journal = {\apj},
         year = 2023,
        month = jun,
       volume = {949},
       number = {2},
          eid = {109},
        pages = {109},
          doi = {10.3847/1538-4357/acc58f},
archivePrefix = {arXiv},
       eprint = {2304.01718},
 primaryClass = {astro-ph.GA},
       adsurl = {https://ui.adsabs.harvard.edu/abs/2023ApJ...949..109L}
}

@ARTICLE{CASA22,
       author = {{CASA Team} and {Bean}, Ben and {Bhatnagar}, Sanjay and {Castro}, Sandra and {Donovan Meyer}, Jennifer and {Emonts}, Bjorn and {Garcia}, Enrique and {Garwood}, Robert and {Golap}, Kumar and {Gonzalez Villalba}, Justo and {Harris}, Pamela and {Hayashi}, Yohei and {Hoskins}, Josh and {Hsieh}, Mingyu and {Jagannathan}, Preshanth and {Kawasaki}, Wataru and {Keimpema}, Aard and {Kettenis}, Mark and {Lopez}, Jorge and {Marvil}, Joshua and {Masters}, Joseph and {McNichols}, Andrew and {Mehringer}, David and {Miel}, Renaud and {Moellenbrock}, George and {Montesino}, Federico and {Nakazato}, Takeshi and {Ott}, Juergen and {Petry}, Dirk and {Pokorny}, Martin and {Raba}, Ryan and {Rau}, Urvashi and {Schiebel}, Darrell and {Schweighart}, Neal and {Sekhar}, Srikrishna and {Shimada}, Kazuhiko and {Small}, Des and {Steeb}, Jan-Willem and {Sugimoto}, Kanako and {Suoranta}, Ville and {Tsutsumi}, Takahiro and {van Bemmel}, Ilse M. and {Verkouter}, Marjolein and {Wells}, Akeem and {Xiong}, Wei and {Szomoru}, Arpad and {Griffith}, Morgan and {Glendenning}, Brian and {Kern}, Jeff},
        title = "{CASA, the Common Astronomy Software Applications for Radio Astronomy}",
      journal = {\pasp},
         year = 2022,
        month = nov,
       volume = {134},
       number = {1041},
          eid = {114501},
        pages = {114501},
          doi = {10.1088/1538-3873/ac9642},
archivePrefix = {arXiv},
       eprint = {2210.02276},
 primaryClass = {astro-ph.IM},
       adsurl = {https://ui.adsabs.harvard.edu/abs/2022PASP..134k4501C}
}

@ARTICLE{LeungBrown77,
       author = {{Leung}, C.~M. and {Brown}, R.~L.},
        title = "{On the interpretation of carbon monoxide self-absorption profiles seen toward embedded stars in dense interstellar clouds.}",
      journal = {\apjl},
         year = 1977,
        month = jun,
       volume = {214},
        pages = {L73-L78},
          doi = {10.1086/182446},
       adsurl = {https://ui.adsabs.harvard.edu/abs/1977ApJ...214L..73L}
}

@ARTICLE{Tafalla98,
       author = {{Tafalla}, M. and {Mardones}, D. and {Myers}, P.~C. and {Caselli}, P. and {Bachiller}, R. and {Benson}, P.~J.},
        title = "{L1544: A Starless Dense Core with Extended Inward Motions}",
      journal = {\apj},
         year = 1998,
        month = sep,
       volume = {504},
       number = {2},
        pages = {900-914},
          doi = {10.1086/306115},
       adsurl = {https://ui.adsabs.harvard.edu/abs/1998ApJ...504..900T}
}

@ARTICLE{LeeMyers11,
       author = {{Lee}, Chang Won and {Myers}, Philip C.},
        title = "{Internal Motions in Starless Dense Cores}",
      journal = {\apj},
         year = 2011,
        month = jun,
       volume = {734},
       number = {1},
          eid = {60},
        pages = {60},
          doi = {10.1088/0004-637X/734/1/60},
archivePrefix = {arXiv},
       eprint = {1104.2950},
 primaryClass = {astro-ph.SR},
       adsurl = {https://ui.adsabs.harvard.edu/abs/2011ApJ...734...60L}
}

@ARTICLE{Lin25,
       author = {{Lin}, Shuting and {Feng}, Siyi and {Sanhueza}, Patricio and {Wang}, Ke and {Zhang}, Zhi-Yu and {Zhang}, Yichen and {Xu}, Fengwei and {Wang}, Junzhi and {Morii}, Kaho and {Liu}, Hauyu Baobab and {Liu}, Sheng-Yuan and {Wang}, Lile and {Li}, Hui and {Tafoya}, Daniel and {Baan}, Willem and {Li}, Shanghuo and {Sabatini}, Giovanni},
        title = "{The ALMA Survey of 70 {\ensuremath{\mu}}m Dark High-mass Clumps in Early Stages (ASHES). XII. Unanchored Forked Stream in the Propagating Path of a Protostellar Outflow}",
      journal = {\apj},
         year = 2025,
        month = sep,
       volume = {990},
       number = {2},
          eid = {229},
        pages = {229},
          doi = {10.3847/1538-4357/adf208},
archivePrefix = {arXiv},
       eprint = {2507.14564},
 primaryClass = {astro-ph.GA},
       adsurl = {https://ui.adsabs.harvard.edu/abs/2025ApJ...990..229L}
}

@ARTICLE{Izumi24,
       author = {{Izumi}, Natsuko and {Sanhueza}, Patricio and {Koch}, Patrick M. and {Lu}, Xing and {Li}, Shanghuo and {Sabatini}, Giovanni and {Olguin}, Fernando A. and {Zhang}, Qizhou and {Nakamura}, Fumitaka and {Tatematsu}, Ken'ichi and {Morii}, Kaho and {Sakai}, Takeshi and {Tafoya}, Daniel},
        title = "{The ALMA Survey of 70 {\ensuremath{\mu}}m Dark High-mass Clumps in Early Stages (ASHES). X. Hot Gas Reveals Deeply Embedded Star Formation}",
      journal = {\apj},
         year = 2024,
        month = mar,
       volume = {963},
       number = {2},
          eid = {163},
        pages = {163},
          doi = {10.3847/1538-4357/ad18c6},
archivePrefix = {arXiv},
       eprint = {2312.03935},
 primaryClass = {astro-ph.GA},
       adsurl = {https://ui.adsabs.harvard.edu/abs/2024ApJ...963..163I}
}

@ARTICLE{Morii26,
       author = {{Morii}, Kaho and {Sanhueza}, Patricio and {Zhang}, Qizhou and {Sabatini}, Giovanni and {Li}, Shanghuo and {Louvet}, Fabien and {Beuther}, Henrik and {Olguin}, Fernando A. and {Lin}, Shuting and {Tafoya}, Daniel and {Sakai}, Takeshi and {Lu}, Xing and {Nakamura}, Fumitaka},
        title = "{The ALMA Survey of 70 {\textmu}m Dark High-mass Clumps in Early Stages (ASHES). XIII. Core Mass Function, Lifetime, and Growth of Prestellar Cores}",
      journal = {\apj},
         year = 2026,
        month = feb,
       volume = {997},
       number = {2},
          eid = {155},
        pages = {155},
          doi = {10.3847/1538-4357/ae25f6},
archivePrefix = {arXiv},
       eprint = {2512.00147},
 primaryClass = {astro-ph.GA},
       adsurl = {https://ui.adsabs.harvard.edu/abs/2026ApJ...997..155M}
}

@software{Kepley19-automask,
       author = {{Kepley}, Amanda A.},
        title = "{Auto-multithresh: Automated masking for clean}",
 howpublished = {Astrophysics Source Code Library, record ascl:1909.001},
         year = 2019,
        month = sep,
          eid = {ascl:1909.001},
archivePrefix = {ascl},
       eprint = {1909.001},
       adsurl = {https://ui.adsabs.harvard.edu/abs/2019ascl.soft09001K}
}

@ARTICLE{VazquezSemadeni24,
       author = {{V{\'a}zquez-Semadeni}, Enrique and {G{\'o}mez}, Gilberto C. and {Gonz{\'a}lez-Samaniego}, Alejandro},
        title = "{Multiscale accretion in dense cloud cores and the delayed formation of massive stars}",
      journal = {\mnras},
         year = 2024,
        month = may,
       volume = {530},
       number = {3},
        pages = {3445-3457},
          doi = {10.1093/mnras/stae1090},
archivePrefix = {arXiv},
       eprint = {2306.13846},
 primaryClass = {astro-ph.GA},
       adsurl = {https://ui.adsabs.harvard.edu/abs/2024MNRAS.530.3445V}
}

@ARTICLE{HoHaschick86,
       author = {{Ho}, P.~T.~P. and {Haschick}, A.~D.},
        title = "{Molecular Clouds Associated with Compact H II Regions. III. Spin-up and Collapse in the Core of G10.6-0.4}",
      journal = {\apj},
         year = 1986,
        month = may,
       volume = {304},
        pages = {501},
          doi = {10.1086/164184},
       adsurl = {https://ui.adsabs.harvard.edu/abs/1986ApJ...304..501H}
}

@ARTICLE{KetoHoHaschick88,
       author = {{Keto}, Eric R. and {Ho}, Paul T.~P. and {Haschick}, Aubrey D.},
        title = "{The Observed Structure of the Accretion Flow around G10.6-0.4}",
      journal = {\apj},
         year = 1988,
        month = jan,
       volume = {324},
        pages = {920},
          doi = {10.1086/165949},
       adsurl = {https://ui.adsabs.harvard.edu/abs/1988ApJ...324..920K}
}

@ARTICLE{Jackson26,
       author = {{Jackson}, James M. and {Whitaker}, J. Scott and {Chambers}, Edward and {Simon}, Robert and {Guevara}, Cristian and {Allingham}, David and {Patterson}, Philippa and {Killerby-Smith}, Nicholas and {Askew}, Jacob and {Sanhueza}, Patricio and {Stephens}, Ian W. and {Schmiedeke}, Anika and {Loughnane}, Robert},
        title = "{Dense Molecular Clumps with Large Blue Asymmetries: Evidence for Collapse}",
      journal = {\apj},
         year = 2026,
        month = feb,
       volume = {998},
       number = {1},
          eid = {167},
        pages = {167},
          doi = {10.3847/1538-4357/ae19eb},
archivePrefix = {arXiv},
       eprint = {2602.13992},
 primaryClass = {astro-ph.GA},
       adsurl = {https://ui.adsabs.harvard.edu/abs/2026ApJ...998..167J}
}

@ARTICLE{ZhangHo97,
       author = {{Zhang}, Qizhou and {Ho}, Paul T.~P.},
        title = "{Dynamical Collapse in W51 Massive Cores: NH$_{3}$ Observations}",
      journal = {\apj},
         year = 1997,
        month = oct,
       volume = {488},
       number = {1},
        pages = {241-257},
          doi = {10.1086/304667},
       adsurl = {https://ui.adsabs.harvard.edu/abs/1997ApJ...488..241Z}
}

@ARTICLE{ZhangHoOhashi98,
       author = {{Zhang}, Qizhou and {Ho}, Paul T.~P. and {Ohashi}, Nagayoshi},
        title = "{Dynamical Collapse in W51 Massive Cores: CS (3-2) and CH$_{3}$CN Observations}",
      journal = {\apj},
         year = 1998,
        month = feb,
       volume = {494},
       number = {2},
        pages = {636-656},
          doi = {10.1086/305243},
       adsurl = {https://ui.adsabs.harvard.edu/abs/1998ApJ...494..636Z}
}

@ARTICLE{FullerWilliamsSridharan05,
       author = {{Fuller}, G.~A. and {Williams}, S.~J. and {Sridharan}, T.~K.},
        title = "{The circumstellar environment of high mass protostellar objects. III. Evidence of infall?}",
      journal = {\aap},
         year = 2005,
        month = nov,
       volume = {442},
       number = {3},
        pages = {949-959},
          doi = {10.1051/0004-6361:20042110},
archivePrefix = {arXiv},
       eprint = {astro-ph/0508098},
 primaryClass = {astro-ph},
       adsurl = {https://ui.adsabs.harvard.edu/abs/2005A&A...442..949F}
}

@ARTICLE{He16,
       author = {{He}, Yu-Xin and {Zhou}, Jian-Jun and {Esimbek}, Jarken and {Ji}, Wei-Guang and {Wu}, Gang and {Tang}, Xin-Di and {Komesh}, Toktarkhan and {Yuan}, Ye and {Li}, Da-Lei and {Baan}, W.~A.},
        title = "{Properties of massive star-forming clumps with infall motions}",
      journal = {\mnras},
         year = 2016,
        month = sep,
       volume = {461},
       number = {3},
        pages = {2288-2308},
          doi = {10.1093/mnras/stw1301},
archivePrefix = {arXiv},
       eprint = {1605.09024},
 primaryClass = {astro-ph.GA},
       adsurl = {https://ui.adsabs.harvard.edu/abs/2016MNRAS.461.2288H}
}
\bibliographystyle{aasjournalv7}

\end{document}